\documentclass[reprint,superscriptaddress,nobibnotes,amsmath,amssymb,twocolumn,prb]{revtex4-1}
\usepackage{bm,graphicx,tabularx,array,dcolumn,xcolor,multirow,amsmath,amssymb,amsfonts,physics,siunitx}
\usepackage[version=4]{mhchem}
\usepackage[utf8]{inputenc}
\usepackage[T1]{fontenc}
\usepackage{txfonts,dsfont}
\usepackage{xspace}

\usepackage[normalem]{ulem}
\newcommand{\be}{\begin{equation}}
\newcommand{\ee}{\end{equation}}

\usepackage[
colorlinks=true,
citecolor=blue,
linkcolor=blue,
filecolor=blue,      
urlcolor=blue,	
breaklinks=true
	]{hyperref}
\newcommand{\latin}[1]{#1}
\newcommand{\ie}{\latin{i.e.}}
\newcommand{\eg}{\latin{e.g.}}

\newcommand{\mcc}[1]{\multicolumn{1}{c}{#1}}

\newcommand{\cP}{\mathcal{P}}
\newcommand{\cS}{\mathcal{S}}

\newcommand{\cC}{\mathcal{C}}
\newcommand{\cD}{\mathcal{D}}
\newcommand{\cE}{\mathcal{E}}
\newcommand{\cI}{\mathcal{I}}
\newcommand{\cH}{\mathcal{H}}

\newcommand{\PsiT}{\Psi_\text{T}}
\newcommand{\PsiG}{\Psi^{+}}
\newcommand{\EL}{E_\text{L}}
\newcommand{\Id}{\mathds{1}}

\newcolumntype{Y}{>{\centering\arraybackslash}X}

\newcommand{\LCPQ}{Laboratoire de Chimie et Physique Quantiques (UMR 5626), Universit\'e de Toulouse, CNRS, UPS, France}
\newcommand{\LCT}{Laboratoire de Chimie Th\'eorique, Sorbonne-Universit\'e, Paris, France}

\begin{document}	
\title{Diffusion Monte Carlo using domains in configuration space}
\author{Roland Assaraf}
        \affiliation{\LCT}
\author{Emmanuel Giner}
        \affiliation{\LCT}
\author{Vijay Gopal Chilkuri}
       \affiliation{\LCPQ}
\author{Pierre-Fran\c{c}ois Loos}
        \affiliation{\LCPQ}
\author{Anthony Scemama}
        \affiliation{\LCPQ}
\author{Michel Caffarel}
\email[Corresponding author: ]{caffarel@irsamc.ups-tlse.fr}
        \affiliation{\LCPQ}
\begin{abstract}

\noindent
The sampling of the configuration space in diffusion Monte Carlo (DMC) is done using walkers moving randomly.
In a previous work on the Hubbard model [\href{https://doi.org/10.1103/PhysRevB.60.2299}{Assaraf et al.~Phys.~Rev.~B \textbf{60}, 2299 (1999)}],
it was shown that the probability for a walker to stay a certain amount of time in the same state obeys a Poisson law and that the on-state dynamics can be integrated out exactly, leading to an effective dynamics connecting only different states.
Here, we extend this idea to the general case of a walker trapped within domains of arbitrary shape and size.
The equations of the resulting effective stochastic dynamics are derived.
The larger the average (trapping) time spent by the walker within the domains, the greater the reduction in statistical fluctuations.
A numerical application to the Hubbard model is presented. 
Although this work presents the method for (discrete) finite linear spaces, it can be generalized without fundamental difficulties to continuous configuration spaces.
\end{abstract}

\maketitle

\section{Introduction}

Diffusion Monte Carlo (DMC) is a class of stochastic methods for evaluating the ground-state properties of quantum systems. 
They have been extensively used in virtually all domains of physics and chemistry where the many-body quantum problem plays a central role (condensed-matter physics,\cite{Foulkes_2001,Kolorenc_2011} quantum liquids,\cite{Holzmann_2006} nuclear physics,\cite{Carlson_2015,Carlson_2007} theoretical chemistry,\cite{Austin_2012} etc). 
DMC can be used either for systems defined in a continuous configuration space (typically, a set of particles moving in space) for which the Hamiltonian operator is defined in a Hilbert space of infinite dimension or systems defined in a discrete configuration space where the Hamiltonian reduces to a matrix. 
Here, we shall consider only the discrete case, that is, the general problem of calculating the lowest eigenvalue and/or eigenstate of a (very large) matrix.
The generalization to continuous configuration spaces presents no fundamental difficulty.

In essence, DMC is based on \textit{stochastic} power methods, a family of well-established numerical approaches able to extract the largest or smallest eigenvalues of a matrix (see, \eg, Ref.~\onlinecite{Golub_2012}). 
These approaches are particularly simple as they merely consist in applying a given matrix (or some simple function of it) as many times as required on some arbitrary vector belonging to the linear space. 
Thus, the basic step of the corresponding algorithm essentially reduces to successive matrix-vector multiplications.
In practice, power methods are employed under more sophisticated implementations, such as, \eg, the Lancz\`os algorithm (based on Krylov subspaces) \cite{Golub_2012} or Davidson's method where a diagonal preconditioning is performed. \cite{Davidson_1975} 
When the size of the matrix is too large, matrix-vector multiplications become unfeasible and probabilistic techniques to sample only the most important contributions of the matrix-vector product are required. 
This is the basic idea of DMC. 
There exist several variants of DMC known under various names: pure DMC, \cite{Caffarel_1988} DMC with branching, \cite{Reynolds_1982} reptation Monte Carlo, \cite{Baroni_1999} stochastic reconfiguration Monte Carlo, \cite{Sorella_1998,Assaraf_2000} etc. 
Here, we shall place ourselves within the framework of pure DMC whose mathematical simplicity is particularly appealing when developing new ideas. 
However, all the ideas presented in this work can be adapted without too much difficulty to the other variants, so the denomination DMC must ultimately be understood here as a generic name for this broad class of methods.

Without entering into the mathematical details (which are presented below), the main ingredient of DMC in order to perform the matrix-vector multiplications probabilistically is to introduce a stochastic matrix (or transition probability matrix) that generates stepwise a series of states over which statistical averages are evaluated.
The critical aspect of any Monte Carlo scheme is the amount of computational effort required to reach a given statistical error.
Two important avenues to decrease the error are the use of variance reduction techniques (for example, by introducing improved estimators \cite{Assaraf_1999A}) or to improve the quality of the sampling (minimization of the correlation time between states).
Another possibility, at the heart of the present work, is to integrate out exactly some parts of the dynamics, thus reducing the number of degrees of freedom and, hence, the amount of statistical fluctuations. 

In previous works,\cite{Assaraf_1999B,Caffarel_2000} it has been shown that the probability for a walker to stay a certain amount of time in the same state obeys a Poisson law and that the on-state dynamics can be integrated out to generate an effective dynamics connecting only different states with some renormalized estimators for the properties.
Numerical applications have shown that the statistical errors can be very significantly decreased.
Here, we extend this idea to the general case where a walker remains a certain amount of time in a finite domain no longer restricted to a single state. 
It is shown how to define an effective stochastic dynamics describing walkers moving from one domain to another. 
The equations of the effective dynamics are derived and a numerical application for a model (one-dimensional) problem is presented. 
In particular, it shows that the statistical convergence of the energy can be greatly enhanced when domains associated with large average trapping times are considered.

It should be noted that the use of domains in quantum Monte Carlo is not new. Domains have been introduced within the context of 
Green's function Monte Carlo (GFMC) pioneered by Kalos \cite{Kalos_1962,Kalos_1970} 
and later developed and applied by Kalos and others. \cite{Kalos_1974,Ceperley_1979,Ceperley_1983,Moskowitz_1986} 
In GFMC, an approximate Green's function that can be sampled is required for the stochastic propagation of the wave function. 
In the so-called domain GFMC version of GFMC introduced in Ref.~\onlinecite{Kalos_1970} and \onlinecite{Kalos_1974} 
the sampling is realized by using the restriction of the Green's function 
to a small domain consisting of the cartesian product of small spheres around each particle, the potential being considered constant within the domain. Fundamentally, the method presented in this work is closely related to the domain GFMC, although the way we present the formalism in terms of walkers
trapped within domains
and derive the equations that may appear different. 
However, we show here how to use 
domains of arbitrary size, a new feature that greatly enhances the efficiency of the simulations when domains are suitably chosen, as illustrated in our numerical 
application.

Finally, from a general perspective, it is interesting to emphasize that the present method illustrates how suitable combinations of stochastic and deterministic techniques lead to a more efficient and valuable method.
In recent years, a number of works have exploited this idea and proposed hybrid stochastic/deterministic schemes.
Let us mention, for example, the semi-stochastic approach of Petruzielo {\it et al.}, \cite{Petruzielo_2012} two different hybrid algorithms for evaluating the second-order perturbation energy in selected configuration interaction methods, \cite{Garniron_2017b,Sharma_2017} the approach of Willow \textit{et al.} for computing stochastically second-order many-body perturbation energies, \cite{Willow_2012} or the zero variance Monte Carlo scheme for evaluating two-electron integrals in quantum chemistry. \cite{Caffarel_2019}

The paper is organized as follows. 
Section \ref{sec:DMC} presents the basic equations and notations of DMC. 
First, the path integral representation of the Green's function is given in Subsec.~\ref{sec:path}. 
Second, the probabilistic framework allowing the Monte Carlo calculation of the Green's function is presented in Subsec.~\ref{sec:proba}. 
Section \ref{sec:DMC_domains} is devoted to the use of domains in DMC. 
We recall in Subsec.~\ref{sec:single_domains} the case of a domain consisting of a single state. 
The general case is then treated in Subsec.~\ref{sec:general_domains}. 
In Subsec.~\ref{sec:Green}, both the time- and energy-dependent Green's function using domains are derived. 
Section \ref{sec:numerical} presents the application of the approach to the one-dimensional Hubbard model. 
Finally, in Sec.\ref{sec:conclu}, some conclusions and perspectives are given.
Atomic units are used throughout.

\section{Diffusion Monte Carlo}
\label{sec:DMC}

\subsection{Path-integral representation}
\label{sec:path}

As previously mentioned, DMC is a stochastic implementation of the power method defined by the following operator:
\be
	T = \Id -\tau (H-E\Id),
\ee 
where $\Id$ is the identity operator, $\tau$ a small positive parameter playing the role of a time step, $E$ some arbitrary reference energy, and $H$ the Hamiltonian operator. For any initial vector $\ket{\Psi_0}$ provided that $\braket{\Phi_0}{\Psi_0} \ne 0$ and for $\tau$ sufficiently small, we have
\be
\label{eq:limTN}
	\lim_{N \to \infty} T^N \ket{\Psi_0} = \ket{\Phi_0},
\ee
where $\ket{\Phi_0}$ is the ground-state wave function, \ie, $H \ket{\Phi_0} = E_0 \ket{\Phi_0}$.
The equality in Eq.~\eqref{eq:limTN} holds up to a global phase factor playing no role in physical quantum averages. 
At large but finite $N$, the vector $T^N \ket{\Psi_0}$ differs from $\ket{\Phi_0}$ only by an exponentially small correction, making it straightforward to extrapolate the finite-$N$ results to $N \to \infty$.

Likewise, ground-state properties may be obtained at large $N$. 
For example, in the important case of the energy, one can project out the vector $T^N \ket{\Psi_0}$ on some approximate vector, $\ket{\PsiT}$, as follows
\be
\label{eq:E0}
	E_0 = \lim_{N \to \infty } \frac{\mel{\Psi_0}{T^N}{H\Psi_T}}{\mel{\Psi_0}{T^N}{\Psi_T}}.
\ee
$\ket{\PsiT}$ is known as the trial wave vector (function), and is chosen as an approximation of the true ground-state vector.

To proceed further we introduce the time-dependent Green's matrix $G^{(N)}$ defined as
\be
	G^{(N)}_{ij}=\mel{i}{T^N}{j}.
\ee
where $\ket{i}$ and $\ket{j}$ are basis vectors.
The denomination ``time-dependent Green's matrix'' is used here since $G$ may be viewed as  a short-time approximation of the (time-imaginary) evolution operator $e^{-N\tau H}$, which is usually referred to as the imaginary-time dependent Green's function.

Introducing the set of $N-1$ intermediate states, $\{ \ket{i_k} \}_{1 \le k \le N-1}$, between each $T$ in $T^N$, $G^{(N)}$ can be written in the following expanded form
\be
\label{eq:cn}
	G^{(N)}_{i_0 i_N} = \sum_{i_1} \sum_{i_2} \cdots \sum_{i_{N-1}} \prod_{k=0}^{N-1} T_{i_{k} i_{k+1}},
\ee
where $T_{ij} =\mel{i}{T}{j}$.
Here, each index $i_k$ runs over all basis vectors.

In quantum physics, Eq.~\eqref{eq:cn} is referred to as the path-integral representation of the Green's matrix (or function).
The series of states $\ket{i_0}, \ldots,\ket{i_N}$ is interpreted as a ``path'' in the Hilbert space starting at vector $\ket{i_0}$ and ending at vector $\ket{i_N}$, where $k$ plays the role of a time index. 
Each path is associated with a weight $\prod_{k=0}^{N-1} T_{i_{k} i_{k+1}}$ and the path integral expression of $G$ can be recast in the more suggestive form as follows:
\be
\label{eq:G}
	G^{(N)}_{i_0 i_N}= \sum_{\text{all paths $\ket{i_1},\ldots,\ket{i_{N-1}}$}} \prod_{k=0}^{N-1} T_{i_{k} i_{k+1}}.
\ee

\begin{figure*}
\includegraphics[width=0.7\textwidth]{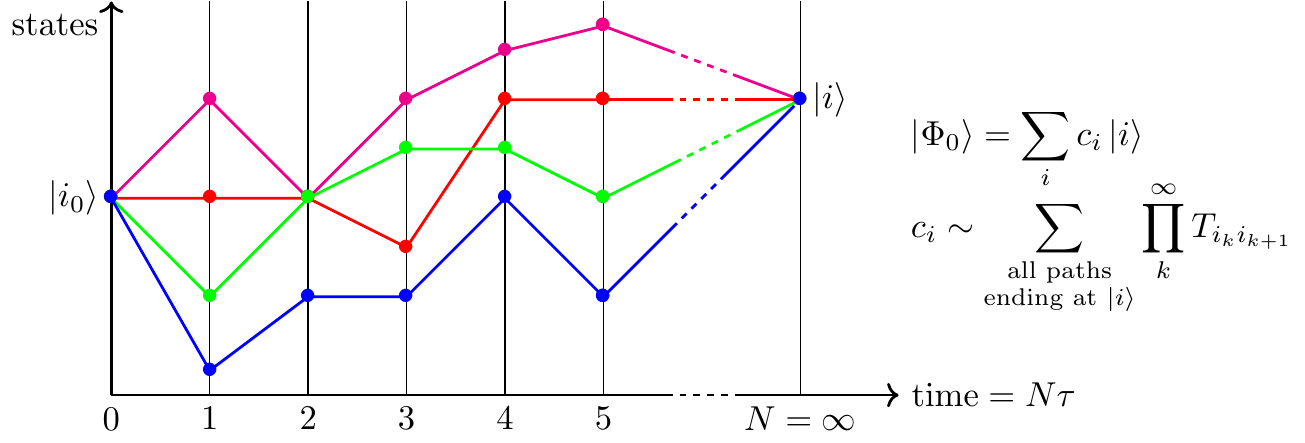}
\caption{
Path integral representation of the exact coefficient $c_i=\braket{i}{\Phi_0}$ of the ground-state wave function $\ket{\Phi_0}$ obtained as an infinite sum of paths starting from $\ket{i_0}$ and ending at $\ket{i}$ [see Eq.\eqref{eq:G}]. 
Each path carries a weight $\prod_k T_{i_{k} i_{k+1}}$ computed along it. 
The result is independent of the choice of the initial state  $\ket{i_0}$, provided that $\braket{i_0}{\Phi_0} \neq 0$.
Here, only four paths of infinite length have been represented.}
\label{fig:paths}
\end{figure*}

This expression allows a simple and vivid interpretation of the solution. 
In the limit $N \to \infty$, the $i_N$th component of the ground-state wave function (obtained as $\lim_{N \to \infty} G^{(N)}_{i_0 i_N})$ is the weighted sum over all possible paths arriving at vector $\ket{i_N}$. 
This result is independent of the initial vector $\ket{i_0}$, apart from some irrelevant global phase factor. 
We illustrate this fundamental property pictorially in Fig.~\ref{fig:paths}.
When the size of the linear space is small, the explicit calculation of the full sums involving $M^N$ terms (where $M$ is the size of the Hilbert space) can be performed. 
In such a case, we are in the realm of what one would call ``deterministic'' power methods, such as the Lancz\`os or Davidson approaches. 
If not, probabilistic techniques for generating only the paths contributing significantly to the sums are to be used.
This is the central theme of the present work.

\subsection{Probabilistic framework}
\label{sec:proba}

To derive a probabilistic expression for the Green's matrix, we introduce a guiding wave function $\ket{\PsiG}$ having strictly positive components, \ie, $\PsiG_i > 0$, in order to perform a similarity transformation of the operators $G^{(N)}$ and $T$, 
\begin{align}
\label{eq:defT}
	\bar{T}_{ij} & = \frac{\PsiG_j}{\PsiG_i} T_{ij},
	&
	\bar{G}^{(N)}_{ij}& = \frac{\PsiG_j}{\PsiG_i} G^{(N)}_{ij}.
\end{align}
Note that, thanks to the properties of similarity transformations, the path integral expression relating $G^{(N)}$ and $T$  [see Eq.~\eqref{eq:G}] remains unchanged for $\bar{G}^{(N)}$ and $\bar{T}$.

Now, the key idea to take advantage of probabilistic techniques is to rewrite the matrix elements of $\bar{T}$ as those of a stochastic matrix multiplied by some residual weights (here, not necessarily positive), namely
\be
\label{eq:defTij}
	\bar{T}_{ij}= p_{i \to j} w_{ij}.
\ee
Here, we recall that a stochastic matrix is defined as a matrix with positive entries that obeys
\be
\label{eq:sumup}
	\sum_j p_{i \to j}=1.
\ee
Using this representation for $\bar{T}_{ij}$ the similarity-transformed Green's matrix components can be rewritten as
\be
\label{eq:GN_simple}
        \bar{G}^{(N)}_{i_0 i_N} =
        \sum_{i_1,\ldots,i_{N-1}} \qty( \prod_{k=0}^{N-1} p_{i_{k} \to i_{k+1}} ) \prod_{k=0}^{N-1} w_{i_{k}i_{k+1}},
\ee
which is amenable to Monte Carlo calculations by generating paths using the transition probability matrix $p_{i \to j}$.

Let us illustrate this in the case of the energy as given by Eq.~\eqref{eq:E0}. Taking $\ket{\Psi_0}=\ket{i_0}$ as initial state, we have
\be
        E_0 = \lim_{N \to \infty }
        \frac{ \sum_{i_N} G^{(N)}_{i_0 i_N} {(H\PsiT)}_{i_N} }
        { \sum_{i_N} {G}^{(N)}_{i_0 i_N} {\PsiT}_{i_N} }.
\ee
which can be rewritten probabilistically as
\be
        E_0 = \lim_{N \to \infty }
        \frac{ \expval{ \prod_{k=0}^{N-1} w_{i_{k}i_{k+1}} \frac{ {(H\PsiT)}_{i_N} }{ \PsiG_{i_N} }}}
        { \expval{ \prod_{k=0}^{N-1} w_{i_{k}i_{k+1}} \frac{ {\PsiT}_{i_N} } {\PsiG_{i_N}} }},
\ee
where $\expval{...}$ is the probabilistic average defined over the set of paths $\ket{i_1},\ldots,\ket{i_N}$ occurring with probability
\be
 \text{Prob}_{i_0}(i_1,\ldots,i_{N}) = \prod_{k=0}^{N-1} p_{i_{k} \to i_{k+1}}.
\ee
Using Eq.~\eqref{eq:sumup} and the fact that $p_{i \to j} \ge 0$, one can easily verify that $\text{Prob}_{i_0}$ is positive and obeys
\be
        \sum_{i_1,\ldots,i_{N}} \text{Prob}_{i_0}(i_1,\ldots,i_{N}) = 1,
\ee
as it should.

The rewriting of $\bar{T}_{ij}$ as a product of a stochastic matrix times some general real weight 
does not introduce any constraint on the choice of the stochastic matrix, so that, in theory, any stochastic matrix could be used.  However, 
in practice, it is highly desirable that the magnitude of the fluctuations of the weight during the Monte Carlo simulation be as small as
possible. A natural solution is to choose a stochastic matrix as close as possible to $\bar{T}_{ij}$. This is done as follows.

Let us introduce the  following operator
\be
\label{eq:T+}
	T^+= \Id - \tau \qty( H^+ - \EL^+ \Id ), 
\ee
where 
$H^+$ is the matrix obtained from $H$ by imposing the off-diagonal elements to be negative
\be
    H^+_{ij}=
    \begin{cases}
    	\phantom{-}H_{ij},	&	\text{if $i=j$},	
    	\\
    	-\abs{H_{ij}},		&	\text{if $i\neq j$}.
    \end{cases}
\ee
Here, $\EL^+ \Id$ is the diagonal matrix whose diagonal elements are defined as
\be
	(\EL^+)_{i}= \frac{\sum_j H^+_{ij}\PsiG_j}{\PsiG_i}.
\ee
The vector $\EL^+$ is known as the local energy vector associated with $\PsiG$.
By construction, the operator $H^+ - \EL^+ \Id$ in the definition of $T^+$ [see Eq.~\eqref{eq:T+}] has been chosen to admit $\ket{\PsiG}$ as a ground-state wave function with zero eigenvalue, \ie, $\qty(H^+ - E_L^+ \Id) \ket{\PsiG} = 0$, leading to the relation
\be
\label{eq:relT+}
	T^+ \ket{\PsiG} = \ket{\PsiG}.
\ee

We are now in the position to define the stochastic matrix as
\be
\label{eq:pij}
	p_{i \to j} 
	= \frac{\PsiG_j}{\PsiG_i} T^+_{ij}
	= 
	\begin{cases}
		1 - \tau \qty[ H^+_{ii}- (\EL^+)_{i} ],					&	\qif* i=j,	
		\\
		\tau \frac{\PsiG_{j}}{\PsiG_{i}} \abs{H_{ij}}  \ge 0, 	&	\qif* i\neq j.
	\end{cases}
\ee
As readily seen in Eq.~\eqref{eq:pij}, the off-diagonal terms of the stochastic matrix are positive, while the diagonal terms can be made positive if $\tau$ is chosen sufficiently small via the condition 
\be
\label{eq:cond}
	\tau \leq \frac{1}{\max_i\abs{H^+_{ii}-(\EL^+)_{i}}}.
\ee
The sum-over-states condition [see Eq.~\eqref{eq:sumup}],
\be
	\sum_j p_{i \to j}= \frac{\mel{i}{T^+}{\PsiG}}{\PsiG_{i}} = 1,
\ee
follows from the fact that $\ket{\PsiG}$ is eigenvector of $T^+$, as evidenced by Eq.~\eqref{eq:relT+}.
This ensures that $p_{i \to j}$ is indeed a stochastic matrix.

At first sight, the condition defining the maximum value of $\tau$ [see Eq.~\eqref{eq:cond}] may appear rather tight since, for very large matrices, it may impose an extremely small value of the time step. 
However, in practice, during the simulation only a (tiny) fraction of the linear space is sampled, and the maximum absolute value of $H^+_{ii}-(\EL^+)_{i}$ for the sampled states turns out to be not too large.
Hence, reasonable values of $\tau$ can be selected without violating the positivity of the transition probability matrix.
Note that one can eschew this condition via a simple generalization of the transition probability matrix:
\be
	p_{i \to j} 
	= \frac{ \frac{\PsiG_{j}}{\PsiG_{i}} \abs{\mel{i}{T^+}{j}} } 
	{ \sum_j \frac{\PsiG_{j}}{\PsiG_{i}} \abs{\mel{i}{T^+}{j}} }
	= \frac{ \PsiG_{j} \abs*{T^+_{ij}} }
	{ \sum_j \PsiG_{j} \abs*{T^+_{ij}} }.
\ee
This new transition probability matrix with positive entries reduces to Eq.~\eqref{eq:pij} when $T^+_{ij}$ is positive as $\sum_j \PsiG_{j} T^+_{ij} = 1$.

Now, we need to make the connection between the transition probability matrix, $p_{i \to j}$, defined from the Hamiltonian $H^{+}$ via $T^+$ and the operator $T$ associated with the exact Hamiltonian $H$. 
This is done thanks to Eq.~\eqref{eq:defTij} that connects $p_{i \to j}$ and $T_{ij}$ through the weight
\be
	w_{ij}=\frac{T_{ij}}{T^+_{ij}},
\ee
derived from Eqs.~\eqref{eq:defT} and \eqref{eq:pij}.

To calculate the probabilistic averages, an artificial (mathematical) ``particle'' called a walker (or psi-particle) is introduced.
During the Monte Carlo simulation, the walker moves in configuration space by drawing new states with 
probability $p_{i_k \to i_{k+1}}$, thus realizing the path of probability  $\text{Prob}_{i_0}$. 
Note that, instead of using a single walker, it is common to introduce a population of independent walkers and to calculate the averages over this population.
In addition, thanks to the ergodicity property of the stochastic matrix (see, for example, Ref.~\onlinecite{Caffarel_1988}), a unique path of infinite length from which sub-paths of length $N$ can be extracted may also be used. 
We shall not insist here on these practical details that are discussed, for example, in Refs.~\onlinecite{Foulkes_2001,Kolorenc_2011}.


\section{Diffusion Monte Carlo with domains}
\label{sec:DMC_domains}

\subsection{Single-state domains}
\label{sec:single_domains}

During the simulation, walkers move from state to state with the possibility of being trapped a certain number of times in the same state before 
exiting to a different state. This fact can be exploited in order to integrate out some parts of the dynamics, thus leading to a reduction of the statistical 
fluctuations. This idea was proposed some time ago and applied to the SU($N$) one-dimensional Hubbard model.\cite{Assaraf_1999A,Assaraf_1999B,Caffarel_2000} 

Considering a given state $\ket{i}$, the probability that a walker remains exactly $n$ times in $\ket{i}$ (with $ n \ge 1 $) and then exits to a different state $j$ (with $j \neq i$) is
\be
	\cP_{i \to j}(n) = \qty(p_{i \to i})^{n-1}  p_{i \to j}.
\ee
Using the relation 
\be
	\sum_{n=1}^{\infty} (p_{i \to i})^{n-1}=\frac{1}{1-p_{i \to i}}
\ee
and the normalization of the $p_{i \to j}$'s [see Eq.~\eqref{eq:sumup}], one can check that the probability is properly normalized, \ie,
\be
	\sum_{j \ne i} \sum_{n=1}^{\infty} \cP_{i \to j}(n) = 1.
\ee

Naturally, the probability of being trapped during $n$ steps is obtained by summing over all possible exit states
\be
P_i(n)=\sum_{j \ne i} \cP_{i \to j}(n) = \qty(p_{i \to i})^{n-1} \qty( 1 - p_{i \to i} ),
\ee
and this defines a Poisson law with an average number of trapping events
\be
	\bar{n}_i= \sum_{n=1}^{\infty} n P_i(n) = \frac{1}{1 -p_{i \to i}}.
\ee
Introducing the continuous time $t_i = n_i \tau$,  the average trapping time is thus given by
\be
	\bar{t_i}= \qty[ H^+_{ii}-(\EL^+)_{i} ]^{-1},
\ee
and, in the limit $\tau \to 0$, the Poisson probability takes the usual form
\be
	P_{i}(t) = \frac{1}{\bar{t}_i} \exp(-\frac{t}{\bar{t}_i}).
\ee
The time-averaged contribution of the on-state weight can then be easily calculated to be
\be
	\bar{w}_i= \sum_{n=1}^{\infty} w^n_{ii} P_i(n)= \frac{T_{ii}}{T^+_{ii}} \frac{1-T^+_{ii}}{1-T_{ii}}
\ee
Details of the implementation of this effective dynamics can be in found in Refs.~\onlinecite{Assaraf_1999B} and \onlinecite{Caffarel_2000}.

\subsection{Multi-state domains}
\label{sec:general_domains}

Let us now extend the results of Sec.~\ref{sec:single_domains} to a general domain. 
To do so, we associate to each state $\ket{i}$ a set of states, called the domain of $\ket{i}$ denoted $\cD_i$, consisting of the state $\ket{i}$ plus a certain number of states. 
No particular constraints on the type of domains are imposed. 
For example, domains associated with different states may be identical, and they may or may not have common states. 
The only important condition is that the set of all domains ensures the ergodicity property of the effective stochastic dynamics, that is, starting from any state, there is a non-zero probability to reach any other state in a finite number of steps. 
In practice, it is not difficult to impose such a condition.

Let us write an arbitrary path of length $N$ as 
\be
	\ket{i_0} \to \ket{i_1} \to \cdots \to \ket{i_N},
\ee
where the successive states are drawn using the transition probability matrix $p_{i \to j}$. 
This path belongs to the set of paths that can be represented as follows
\be
\label{eq:eff_series}
	(\ket*{I_0},n_0) \to  (\ket*{I_1},n_1) \to \cdots \to (\ket*{I_p},n_p),
\ee
where $\ket{I_0}=\ket{i_0}$ is the initial state, $n_0$ is the number of times the walker remains in the domain of $\ket{i_0}$ (with $1 \le n_0 \le N+1$), $\ket{I_1}$ is the first exit state that does not belong to $\cD_{i_0}$, $n_1$ is the number of times the walker remains in $\cD_{i_1}$ (with $1 \le n_1 \le N+1-n_0$), $\ket{I_2}$ is the second exit state, and so on.
Here, the integer $p$ (with $0 \le p \le N$) indicates the number of exit events occurring along the path. 
The two extreme values, $p=0$ and $p=N$, correspond to the cases where the walker remains in the initial domain during the entire path, and where the walker exits a domain at each step, respectively. 
In what follows, we shall systematically label exit states with upper-case letters $\ket{I_k}$, while lower-case letters denote elementary states $\ket{i_k}$.
Making this distinction is important since the effective stochastic dynamics used in practical Monte Carlo calculations only involve exit states $\ket{I_k}$, the contribution from the elementary states $\ket{i_k}$ being exactly integrated out.

\begin{figure}
\includegraphics[width=\columnwidth,angle=0]{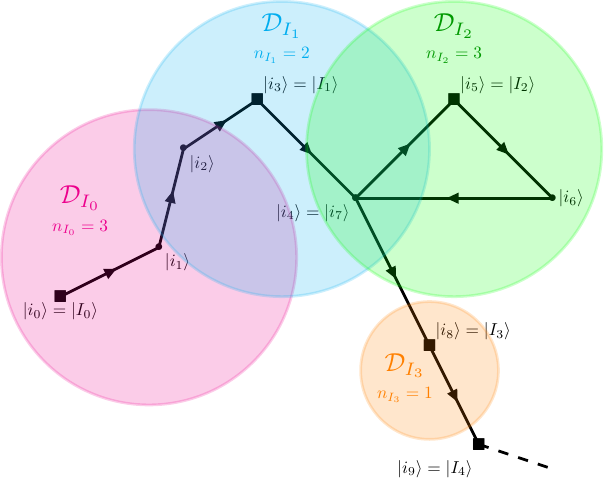}
\caption{Representation of a path in terms of exit states $\ket{I_k}$ and trapping times $\ket{n_k}$. The 
states $\ket{i_k}$ along the path are represented by small black circles and the exit states, $\ket{I_k}$, by larger black squares. 
By convention, the initial state is denoted using a capital letter, \ie,  $\ket{i_0} = \ket{I_0}$, since it is the first state of the effective dynamics involving only exit states. 
See text for additional comments on the time evolution of the path.}
\label{fig:domains}
\end{figure}

Figure \ref{fig:domains} exemplifies how a path can be decomposed as proposed in Eq.~\eqref{eq:eff_series}.
To make things as clear as possible, let us describe how the path drawn in Fig.~\ref{fig:domains} evolves in time.
The walker starts at $\ket{i_0}$ in the domain $\cD_{i_0}$. 
Then, it performs two steps from $\ket{i_0}$ to $\ket{i_2}$ and has left the domain at $\ket{i_3}$, which is thus the first exit state, \ie, $\ket{I_1} = \ket{i_3}$. 
The trapping time in $\cD_{i_0}$ is $n_0=3$ since three states in $\cD_{i_0}$ have been visited (namely, $\ket{i_0}$, $\ket{i_1}$, and $\ket{i_2}$).
During the next steps, the domains $\cD_{I_1}$, $\cD_{I_2}$, and $\cD_{I_3}$ are successively visited with $n_1=2$, $n_2=3$, and $n_3=1$, respectively. 
The corresponding exit states are $\ket{I_2}=\ket{i_5}$, $\ket{I_3}=\ket{i_8}$, and $\ket{I_4}=\ket{i_9}$, respectively. 


Generalizing the single-state case treated previously, let us define the probability of remaining $n$ times in the domain of $\ket{I_0}$ and to exit at $\ket{I} \notin \cD_{I_0}$
\be
\label{eq:eq1C}
	\cP_{I_0 \to I}(n) 
	= \sum_{\ket{i_1} \in \cD_{I_0}} \cdots \sum_{\ket{i_{n-1}} \in \cD_{I_0}} 
	p_{I_0 \to i_1} \ldots p_{i_{n-2} \to i_{n-1}} p_{i_{n-1} \to I}.
\ee
Since the sums are restricted to states belonging to the domain, it is convenient to introduce a projector over each domain
\be
\label{eq:pi}
	P_I = \sum_{\ket{i} \in \cD_I} \dyad{i}{i},
\ee
as well as the projection of $T^+$ over $\cD_I$
\be
T^+_I = P_I T^+ P_I, 
\ee
which governs the dynamics of the walkers trapped in this domain.
Using Eqs.~\eqref{eq:pij} and \eqref{eq:eq1C}, the probability can be rewritten as
\be
\label{eq:eq3C}
	\cP_{I_0 \to I}(n) = \frac{1}{\PsiG_{I_0}} \mel{I_0}{\qty(T^+_{I_0})^{n-1} F^+_{I_0}}{I} \PsiG_{I},
\ee
where the operator $F^+_I =  P_I T^+ (1-P_I)$, corresponding to the last move connecting the inside and outside regions of the domain, has the following matrix elements:
 \be
	(F^+_I)_{ij} = 
	\begin{cases}
		T^+_{ij},	&	\qif* \ket{i} \in \cD_{I} \;\;{\rm and}\;\; \ket{j} \notin \cD_{I}, 
		\\
		0,			&	\text{otherwise}.
	\end{cases}
\ee
Physically, $F$ may be seen as a flux operator through the boundary of $\cD_{I}$.

Knowing the probability of remaining $n$ times in the domain and, then, exiting to some state, it is possible to obtain
the probability of being trapped $n$ times in $\cD_{I}$, just by summing over all possible exit states:
\be
\label{eq:PiN}
	P_{I}(n) = \frac{1}{\PsiG_{I}} \mel{ I }{ \qty(T^+_{I})^{n-1} F^+_{I} }{ \PsiG }.
\ee
The normalization of this probability can be verified using the fact that 
\be
\label{eq:relation}
	\qty(T^+_{I})^{n-1} F^+_I = \qty(T^+_{I})^{n-1} T^+ - \qty(T^+_I)^n,
\ee
leading to\footnote{
The property results from the fact that the series is a telescoping series and that the general term
$\mel{ I }{ \qty(T^+_{I})^{n} }{ \PsiG }$ goes to zero as $n$ goes to infinity.}
\be
	\sum_{n=0}^{\infty} P_{I}(n) 
	= \frac{1}{\PsiG_{I}}  \sum_{n=1}^{\infty} \qty[ \mel{ I }{ \qty(T^+_{I})^{n-1} }{ \PsiG }
	-  \mel{ I }{ \qty(T^+_{I})^{n} }{ \PsiG } ] = 1.
\ee
The average trapping time defined as ${\bar t}_{I}={\bar n}_{I} \tau$ where $ {\bar n}_{I}=\sum_n n P_{I}(n)$  is calculated to be
\be 
	{\bar t}_{I}=\frac{1}{\PsiG_I}  \mel{I}{ { \qty[ P_I \qty( H^+ - \EL^+ \Id ) P_I ] }^{-1} }{ \PsiG }. 
\ee
In practice, the various quantities restricted to the domain will be computed by inverting the matrix $(H^+-\EL^+ \Id)$ in $\cD_{I}$. 
Note that it is possible only if the dimension of the domains is not too large (say, less than a few thousand).

\subsection{Time-dependent Green's matrix using domains}
\label{sec:Green}

In this section, we generalize the path-integral expression of the Green's matrix, Eq.~\eqref{eq:G}, to the case where domains are used.
To do so, we need to introduce the Green's matrix restricted to each domain as follows:
\be
	G^{(N),\cD}_{ij}= \mel{ i }{ T_i^N }{ j }.
\ee
where $T_i$ is the projection of the operator $T$ over the domain $\cD_i$ of $\ket{i}$
\be
T_i = P_i T^+ P_i,
\ee
and $P_i$ is the projector, $P_i=\sum_{\ket{k} \in \cD_i} \dyad{k}{k}$, Eq.~\eqref{eq:pi}.

Using the representation of the paths in terms of exit states and trapping times [see Eq.~\eqref{eq:eff_series}], 
the set of paths can be partitioned according to the number $p$ of exit states.
As already noted above, the number of exit states ranges from $p=0$ (no exit of the initial domain) to $p=N$ (exit of the current domain at each time step).
We shall denote $\cC_p$ the set of paths with $p$ exit states.
The time-dependent Green's matrix, Eq.~\eqref{eq:cn}, is then rewritten as
\be
	G^{(N)}_{i_0 i_N} = \sum_{p=0}^N 
	\sum_{\cC_p} \sum_{(i_1,...,i_{N-1}) \in \cC_p}
	\prod_{k=0}^{N-1} \mel{ i_k }{ T }{ i_{k+1} } .
\ee
$\cC_p$ can be partitioned further by considering the subset of paths, denoted by $\cC_p^{I,n}$, 
having $(\ket{I_k}, n_k)$, for $0 \le k \le p$, as exit states and trapping times.
We recall that, by definition of the exit states, $\ket{I_k} \notin \cD_{I_{k-1}}$. The total time must be conserved, so the relation $\sum_{k=0}^p n_k= N+1$ 
must be fulfilled.
Now, the contribution of $\cC_p^{I,n}$ to the path integral is obtained by summing over all elementary paths $(i_1,...,i_{N-1})$ 
of this set, thus giving
\begin{multline}
\sum_{(i_1,...,i_{N-1}) \in \cC_p^{I,n}} \prod_{k=0}^{N-1} \mel{ i_k }{ T }{ i_{k+1} }
\\
= \qty[ \prod_{k=0}^{p-1} \mel{ I_k }{ \qty(T_{I_k})^{n_k-1} F_{I_k} }{ I_{k+1} } ]
         \mel{ I_k }{ \qty(T_{I_p})^{n_p} }{ i_N }
\end{multline}
Finally, collecting all contributions the Green's matrix is written as
\begin{multline}
\label{eq:Gt}
	G^{(N)}_{i_0 i_N}= G^{(N),\cD}_{i_0 i_N} +
	\sum_{p=1}^{N}
	\sum_{\ket{I_1} \notin \cD_{I_0}, \ldots , \ket{I_p} \notin \cD_{I_{p-1}}  }
	\sum_{n_0 \ge 1} \cdots \sum_{n_p \ge 1}
	\delta_{\sum_{k=0}^p n_k,N+1} 
	\\
	\times 
	\qty[ \prod_{k=0}^{p-1} \mel{ I_k }{ \qty(T_{I_k})^{n_k-1} F_{I_k} }{ I_{k+1} } ]
	G^{(n_p-1),\cD}_{I_p i_N},
\end{multline}
where $\delta_{i,j}$ is a Kronecker delta. 
This expression is the path-integral representation of the Green's matrix using only the variables $(\ket{I_k},n_k)$ of the effective dynamics defined over the set of domains.
The standard formula for $G^{(N)}_{i_0 i_N}$ derived above [see Eq.~\eqref{eq:G}] may be considered as the particular case where the walker exits of the current state $\ket{i_k}$ at each step (no domains are introduced), leading to a number of exit events $p$ equal to $N$. 
In this case, we have $\ket{I_k}=\ket{i_k}$, $n_k=1$ (for $0 \le k \le N$), and we are left only with the $p$th component of the sum, that is, $G^{(N)}_{i_0 i_N}=\prod_{k=0}^{N-1} \mel{ I_k }{ F_{I_k} }{ I_{k+1} }$, with $F=T$, thus recovering Eq.~\eqref{eq:G}.
Note that the first contribution $G^{(N),\cD}_{i_0 i_N}$ corresponds to the case $p=0$ and collects all contributions to the Green's matrix coming from paths remaining indefinitely in the domain of $\ket{i_0}$ (no exit event). 
This contribution is isolated from the sum in Eq.~\eqref{eq:Gt} since, as a domain Green's matrix, it is calculated exactly and is not subject to a stochastic treatment.
Note also that the last state $\ket{i_N}$ is never an exit state because of the very definition of our path representation.

In order to compute $G^{(N)}_{i_0 i_N}$ by resorting to  Monte Carlo techniques, let us reformulate Eq.~\eqref{eq:Gt} using the transition probability $\cP_{I \to J}(n)$ introduced in Eq.~\eqref{eq:eq3C}. 
We first rewrite Eq.~\eqref{eq:Gt} under the form
\begin{multline}
	{G}^{(N)}_{i_0 i_N}={G}^{(N),\cD}_{i_0 i_N} + {\PsiG_{i_0}}
	\sum_{p=1}^{N}
	\sum_{\ket{I_1} \notin \cD_{I_0}, \ldots , \ket{I_p} \notin \cD_{I_{p-1}}}
	\sum_{n_0 \ge 1} \cdots \sum_{n_p \ge 1}
	\\
	\times
	\delta_{\sum_{k=0}^p n_k,N+1} \qty{ \prod_{k=0}^{p-1} \qty[ \frac{\PsiG_{I_{k+1}}}{\PsiG_{I_k}} \mel{ I_k }{ \qty(T_{I_k})^{n_k-1} F_{I_k} }{ I_{k+1} 	} ] }
	\frac{{G}^{(n_p-1),\cD}_{I_p i_N}}{\PsiG_{I_p}}.
\end{multline}
Introducing the weights 
\be
	W_{I_k I_{k+1}} = \frac{\mel{ I_k }{ \qty(T_{I_k})^{n_k-1} F_{I_k} }{ I_{k+1} }}{\mel{ I_k }{ \qty(T^{+}_{I_k})^{n_k-1} F^+_{I_k} }{ I_{k+1} }},
\ee
and using the effective transition probability, Eq.~\eqref{eq:eq3C}, we get
\begin{multline}
\label{eq:Gbart}
	{G}^{(N)}_{i_0 i_N} = {G}^{(N),\cD}_{i_0 i_N} 
\\
+ {\PsiG_{i_0}}\sum_{p=1}^{N} \sum_{{(I,n)}_{p,N}} 
	{
	\qty( \prod_{k=0}^{p-1} W_{I_k I_{k+1}} ) \qty( \prod_{k=0}^{p-1}\cP_{I_k \to I_{k+1}}(n_k)  )
	 \frac{1}{\PsiG_{I_p}} {G}^{(n_p-1), \cD}_{I_p i_N} },
\end{multline}
where, for clarity, $\sum_{{(I,n)}_{p,N}}$ is used as a short-hand notation for $\sum_{\ket{I_1} \notin \cD_{I_0}, \ldots , \ket{I_p} \notin \cD_{I_{p-1}}} \sum_{n_0 \ge 1} \cdots \sum_{n_p \ge 1}$ with the constraint $\sum_{k=0}^p n_k=N+1$.

Under this form, ${G}^{(N)}_{i_0 i_N}$ is now amenable to Monte Carlo calculations 
by generating paths using the transition probability matrix $\cP_{I \to J}(n)$. 
For example, in the case of the energy, we start from
\be
        E_0 = \lim_{N \to \infty }
        \frac{ \sum_{i_N} {G}^{(N)}_{i_0 i_N} {(H\PsiT)}_{i_N} }
        { \sum_{i_N} {G}^{(N)}_{i_0 i_N}  {\PsiT}_{i_N} },
\ee
which can be rewritten probabilistically as
\be
        E_0 = \lim_{N \to \infty }
        \frac{ {G}^{(N),\cD}_{i_0 i_N} + {\PsiG_{i_0}}  \sum_{p=1}^{N}  \expval{ \qty( \prod_{k=0}^{p-1} W_{I_k I_{k+1}} ) {\cal H}_{n_p,I_p} }_p}
             { {G}^{(N),\cD}_{i_0 i_N} + {\PsiG_{i_0}} \sum_{p=1}^{N}  \expval{ \qty( \prod_{k=0}^{p-1} W_{I_k I_{k+1}} ) {\cal S}_{n_p,I_p} }_p},
\ee
where $\expval{\cdots}_p$ is the probabilistic average defined over the set of paths with $p$ exit events of probability $\prod_{k=0}^{p-1} \cP_{I_k \to I_{k+1}}(n_k)$, and 
\begin{align}
	\cH_{n_p,I_p} & = \frac{1}{\PsiG_{I_p}} \sum_{i_N} {G}^{(n_p-1),\cD}_{I_p i_N} (H \Psi_T)_{i_N},
	\\
	\cS_{n_p,I_p} & = \frac{1}{\PsiG_{I_p}} \sum_{i_N} {G}^{(n_p-1), \cD}_{I_p i_N} (\Psi_T)_{i_N},
\end{align}
two quantities taking into account the contribution of the trial wave function at the end of the path.

In practice, the present Monte Carlo algorithm is a simple generalization of the standard diffusion Monte Carlo algorithm.
Stochastic paths starting at $\ket{I_0}$ are generated using the probability $\cP_{I_k \to I_{k+1}}(n_k)$ and are stopped when $\sum_k n_k$ is greater than some target value $N$. 
Averages of the weight products $ \prod_{k=0}^{p-1} W_{I_k I_{k+1}} $ times the end-point contributions ${(\cH/\cS)}_{n_p,I_p}$ are computed for each $p$. 
The generation of the paths can be performed using a two-step process. 
First, the integer $n_k$ is drawn using the probability $P_{I_k}(n)$ [see Eq.~\eqref{eq:PiN}] and, then, the exit state $\ket{I_{k+1}}$ is drawn using the conditional probability $\cP_{I_k \to I_{k+1}}(n_k)/P_{I_k}(n_k)$.
%
\subsection{Domain Green's Function Monte Carlo}
\label{sec:energy}

The aim of this section is to show that it is possible to go further by integrating out the trapping times, $n_k$, of the preceding expressions, thus defining a new effective stochastic dynamics involving now only the exit states. 
Physically, it means that we are going to compute exactly the time evolution of all stochastic paths trapped in each domain. 
We shall present two different ways to derive the new dynamics and renormalized probabilistic averages. 
The first one, called the pedestrian way, consists in starting from the preceding time expression for $G$ and making the explicit integration over the $n_k$'s. 
The second, more direct and elegant, is based on the Dyson equation.

\subsubsection{The pedestrian way} 
Let us define the energy-dependent Green's matrix
\be
	G^E_{ij}= \tau \sum_{N=0}^{\infty} \mel{ i }{ T^N }{ j} = \mel{i}{ \qty( H-E \Id )^{-1} }{j}.
\ee
The denomination ``energy-dependent'' is chosen here since 
this quantity is the discrete version of the Laplace transform of the time-dependent Green's function in a continuous space,
usually known under this name.\footnote{As $\tau \to 0$ and $N \to \infty$ with $N\tau=t$, the operator $T^N$ converges to $e^{-t(H-E \Id)}$. We then have $G^E_{ij} \to \int_0^{\infty} dt \mel{i}{e^{-t(H-E \Id)}}{j}$, which is the Laplace transform of the time-dependent Green's function $\mel{i}{e^{-t(H-E \Id)}}{j}$.}
The remarkable property is that thanks to the summation over $N$ up to infinity, the constrained multiple sums appearing in Eq.~\eqref{eq:Gt} can be factorized in terms of a product of unconstrained sums, as follows
\begin{multline}
	\sum_{N=1}^\infty \sum_{p=1}^N \sum_{n_0 \ge 1} \cdots \sum_{n_p \ge 1} \delta_{\sum_{k=0}^p n_k,N+1} F(n_0,\ldots,n_N)
	\\
	= \sum_{p=1}^{\infty} \sum_{n_0=1}^{\infty} \cdots \sum_{n_p=1}^{\infty} F(n_0,\ldots,n_N),
\end{multline}
where $F$ is some arbitrary function of the trapping times.
Using the fact that $G^E_{ij}= \tau \sum_{N=0}^{\infty} G^{(N)}_{ij}$, where $G^{(N)}_{ij}$ is given by Eq.~\eqref{eq:Gt}, and summing over the variables $n_k$, we get
\begin{multline}
\label{eq:eqfond}
  {G}^{E}_{i_0 i_N}
	= {G}^{E,\cD}_{i_0 i_N}
	+ \sum_{p=1}^{\infty} \sum_{I_1 \notin \cD_0, \hdots , I_p \notin \cD_{p-1}} \\
	\qty[ \prod_{k=0}^{p-1} \mel{ I_k }{ {\qty[ P_k \qty( H-E \Id ) P_k ] }^{-1} (-H)(\Id-P_k) }{ I_{k+1} } ]
       {G}^{E,\cD}_{I_p i_N},
\end{multline}
where, ${G}^{E,\cD}$ is the energy-dependent domain's Green matrix defined as ${G}^{E,\cD}_{ij} = \tau \sum_{N=0}^{\infty} \mel{ i }{ T^N_i }{ j}$.

As a didactical example, Appendix \ref{app:A} reports the exact derivation of this formula in the case of a two-state system.

\subsubsection{Dyson equation} 

In fact, there is a more direct way to derive the same equation by resorting to the Dyson equation. Starting from the well-known equality
\be
	\qty(H-E\Id)^{-1} = \qty(H_0-E\Id)^{-1} + \qty(H_0-E\Id)^{-1} (H_0-H) \qty(H-E\Id)^{-1},
\ee
where $H_0$ is some arbitrary reference Hamiltonian, we have the Dyson equation
\be
\label{eq:GE}
	G^E_{ij} = G^E_{0,ij} + \sum_{kl} G^{E}_{0,ik} (H_0-H)_{kl} G^E_{lj},
\ee
with $G^E_{0,ij} = \mel{i}{ \qty( H_0-E \Id )^{-1} }{j}$.
Let us choose $H_0$ such that $\mel{ i }{ H_0 }{ j } = \mel{ i }{ P_i H P_i }{ j }$ for all $i$ and $j$.
Then, the Dyson equation \eqref{eq:GE} becomes
\be
{G}^{E}_{ij}
	= {G}^{E,\cD}_{ij}
	+ \sum_k  \mel{ i }{ {\qty[P_i \qty(H-E \Id)P_i ]}^{-1} (H_0-H) }{ k } {G}^{E}_{kj}.
\ee
Using the following identity
\be
\begin{split}
{\qty[	P_i \qty(H-E \Id) P_i ]}^{-1} (H_0-H) 
	& = { \qty[ P_i \qty(H-E \Id)P_i ]}^{-1} (P_i H P_i - H) 
	\\
	& = {\qty[ P_i \qty(H-E \Id) P_i]}^{-1} (-H) ( \Id -P_i),
\end{split}
\ee
the Dyson equation may be written under the form
\be
 G^E_{ij} =  {G}^{E,\cD}_{ij}
	+ \sum_{k \notin \cD_i}  \mel{ i }{ {\qty[P_i \qty(H-E \Id) P_i]}^{-1}  (-H)( \Id -P_i) }{ k }  {G}^{E}_{kj},
\ee
which is identical to Eq.~\eqref{eq:eqfond} when $G^E_{ij}$ is expanded iteratively.
Let us use as effective transition probability density
\be
	P_{I \to J} = \frac{1} {\PsiG(I)} \mel{ I }{ {\qty[ P_I \qty(H^+ - \EL^+ \Id) P_I]}^{-1} (-H^+) (\Id -P_I) }{ J } \PsiG(J),
\ee
and the weight
\be
	W^E_{IJ} =
	\frac{ \mel{ I }{ {\qty[ P_I \qty(H-E \Id) P_I]}^{-1} (-H)( \Id -P_I) }{ J} }
	{\mel{ I }{ {\qty[ P_I \qty(H^+ - \EL^+ \Id) P_I]}^{-1} (-H^+)( \Id -P_I) }{ J} }.
\ee
Using Eqs.~\eqref{eq:eq1C}, \eqref{eq:eq3C} and \eqref{eq:relation}, one can easily verify that $P_{I \to J} \ge 0$ and $\sum_J P_{I \to J}=1$. 
Finally, the probabilistic expression reads
\be
\label{eq:final_E}
G^E_{i_0 i_N}= {G}^{E,\cD}_{i_0 i_N}
	+ \sum_{p=1}^{\infty} \PsiG_{i_0} \expval{ \qty( \prod_{k=0}^{p-1} W^E_{I_k I_{k+1}} ) 
\frac{{G}^{E,\cD}_{I_p i_N}} { \PsiG_{I_p}} }.
\ee

\subsubsection{Energy estimator} 
To calculate the energy, we introduce the following estimator 
\be
	\cE(E) = \frac{ \mel{ I_0 }{ \qty(H-E \Id)^{-1} }{ H\PsiT } } {\mel{ I_0 }{ \qty(H-E \Id)^{-1} }{ \PsiT } },
\ee
which is the counterpart of the quantity ${\cal E}_N =\frac{ \mel{ I_0 }{ T^N }{ H\PsiT } } {\mel{ I_0 }{T^N}{ \PsiT } }$ used
in the time-dependent formalism. In this case, the energy was easily obtained by taking the large $N$-limit of ${\cal E}_N$, see Eq.(\ref{eq:E0}).
Here, the situation is not as simple and we must find a way to extract the energy from $\cE(E)$.

Using the spectral decomposition of $H$, we have
\be
\label{eq:calE}
        \cE(E) = \frac{ \sum_i \frac{E_i c_i}{E_i-E}}{\sum_i \frac{c_i}{E_i-E}},
\ee
where $c_i = \braket{ I_0 }{ \Phi_i } \braket{ \Phi_i}{  \PsiT }$ and $\Phi_i$ are the eigenstates of $H$, \ie, $H \ket{\Phi_i} = E_i \ket{\Phi_i}$.
The important observation is that for all eigenstates we have $\cE(E_i)= E_i$. Thus, to get the ground-state energy we propose to search for the solution 
of the non-linear equation $\cE(E)= E$ in the vicinity of $E_0$.

In practical Monte Carlo calculations, the DMC energy is obtained by computing a finite number of components $H_p$ and $S_p$ defined as follows
\be
\cE^\text{DMC}(E,p_{max})= \frac{ H_0+ \sum_{p=1}^{p_\text{max}} H^\text{DMC}_p }
                         {S_0 +\sum_{p=1}^{p_\text{max}} S^\text{DMC}_p }.
\ee
For $ p\ge 1$, Eq.~\eqref{eq:final_E} gives
\begin{align}
        H^\text{DMC}_p & = \PsiG_{i_0}\expval{ \qty(\prod_{k=0}^{p-1} W^E_{I_k I_{k+1}}) {\cal H}_{I_p} },
\label{eq:defHp}
        \\
        S^\text{DMC}_p & = \PsiG_{i_0} \expval{ \qty(\prod_{k=0}^{p-1} W^E_{I_k I_{k+1}}) {\cal S}_{I_p} },
\label{eq:defSp}
\end{align}
where
\begin{align}
	\cH_{I_p} & = \frac{1}{\PsiG_{I_p}} \sum_{i_N} {G}^{E,\cD}_{I_p i_N} (H \Psi_T)_{i_N},
	\\
	\cS_{I_p} & = \frac{1}{\PsiG_{I_p}} \sum_{i_N} {G}^{E, \cD}_{I_p i_N} (\Psi_T)_{i_N}.
\end{align}

For $p=0$, the two components are exactly evaluated as
\begin{align}
        H_0 & = \mel{ I_0 }{ {\qty[ P_{I_0} \qty(H-E \Id) P_{I_0} ]}^{-1} }{ H\PsiT },
        \\
        S_0 & = \mel{ I_0 }{ {\qty[ P_{I_0} \qty(H-E \Id) P_{I_0} ]}^{-1} }{ \PsiT }.
\end{align}
Note that the evaluation of $(H_0,S_0)$ is possible as long as the size of the domain is small enough to allow the calculation of the inverse matrix.
Avoiding the stochastic calculation of quantities, such as $H_0$ or $S_0$, that can be evaluated analytically is computationally very appealing as the statistical error associated with these quantities is completely removed. 
We thus suggest extending the exact calculation of $H_p$ and $S_p$ to higher $p$ values, up to the point where the exponential increase of the number of intermediate states involved in the explicit sums makes the calculation unfeasible.

Finally, $\cE^\text{DMC}(E,p_\text{ex},p_\text{max})$ is evaluated in practice as follows
\be
\cE^\text{DMC}(E,p_\text{ex},p_\text{max})= \frac{   \sum_{p=0}^{p_\text{ex}-1} H_p +\sum_{p=p_\text{ex}}^{p_\text{max}} H^\text{DMC}_p }
                         {                 \sum_{p=0}^{p_\text{ex}-1} S_p +\sum_{p=p_\text{ex}}^{p_\text{max}} S^\text{DMC}_p },
\label{eq:E_pex}
\ee
where $p_\text{ex}$ is the number of components computed exactly. 
Let us emphasize that, since the magnitude of $H_p$ and $S_p$ decreases as a function of $p$, most of the statistical error is removed by computing the dominant terms analytically.
This will be illustrated in the numerical application presented below.

It is easy to check that, in the vicinity of the exact energy $E_0$, $\cE(E)$ is a linear function of $E - E_0$.
Therefore, in practice, we compute the value of $\cE(E)$ for several values of $E$, and fit these data using a linear, quadratic or a more complicated function of $E$ in order to obtain, via extrapolation, an estimate of $E_0$. 
In order to have a precise extrapolation of the energy, it is best to compute $\cE(E)$ for values of $E$ as close as possible to the exact energy. 
However, as $E \to E_0$, both the numerators and denominators of Eq.~\eqref{eq:calE} diverge. 
This is reflected by the fact that one needs to compute more and more $p$-components with an important increase in statistical fluctuations.
Thus, from a practical point of view, a trade-off has to be found between the quality of the extrapolation and the amount of statistical fluctuations. 

\section{Numerical application to the Hubbard model}
\label{sec:numerical}

\subsection{Hamiltonian and trial wave function}

Let us consider the one-dimensional Hubbard Hamiltonian for a chain of $N$ sites 
\be
	H= -t \sum_{\expval{ i j } \sigma} \hat{a}^+_{i\sigma} \hat{a}_{j\sigma} 
	+ U \sum_i \hat{n}_{i\uparrow} \hat{n}_{i\downarrow},
\ee
where $\expval{ i j }$ denotes the summation over two neighboring sites, $\hat{a}^+_{i\sigma}$ ($\hat{a}_{i\sigma}$) is the fermionic creation (annihilation) operator of a spin-$\sigma$ electron (with $\sigma$ = $\uparrow$ or $\downarrow$) on site $i$, $\hat{n}_{i\sigma} = \hat{a}^+_{i\sigma} \hat{a}_{i\sigma}$ the number operator, $t$ the hopping amplitude, and $U$ the on-site Coulomb repulsion.
We consider a chain with an even number of sites and open boundary conditions at half-filling, that is, $N_{\uparrow}=N_{\downarrow}=N/2$.
In the site representation, a general vector of the Hilbert space can be written as
\be
	\ket{n} = \ket{n_{1 \uparrow},\ldots,n_{N \uparrow},n_{1 \downarrow},\ldots,n_{N \downarrow}},
\ee
where $n_{i \sigma}=0$ or $1$ is the number of electrons of spin $\sigma$ on site $i$.

For the one-dimensional Hubbard model with open boundary conditions, the components of the ground-state vector have the same sign (say, $c_i > 0$). 
It is then possible to equate the guiding and trial vectors, \ie, $\ket{c^+} = \ket{c_\text{T}}$.
As a trial wave function, we shall employ a generalization of the Gutzwiller wave function \cite{Gutzwiller_1963}
\be
	\braket{ n }{ c_\text{T} }= e^{-\alpha n_D(n)-\beta n_A(n)},
\ee
where $n_D(n)$ is the number of doubly occupied sites for the configuration $\ket{n}$ and $n_A(n)$ the number of nearest-neighbor antiparallel pairs defined as
\be
	n_A(n)= \sum_{\expval{ i j }} n_{i\uparrow} n _{j\downarrow} \delta_{n_{i\uparrow},1} \delta_{n_{j\downarrow},1}.
\ee
The parameters $\alpha$ and $\beta$ are optimized by minimizing the variational energy of the trial wave function, \ie, 
\be
	E_\text{v}(\alpha,\beta) = \frac{\mel{ c_\text{T} }{H }{c_\text{T}}} {\braket{ c_\text{T} }{ c_\text{T} }}.
\ee

\subsection{Domains}

As discussed above, the efficiency of the method depends on the choice of states forming each domain.
As a general guiding principle, it is advantageous to build domains associated with a large average trapping time in order to integrate out the most important part of the Green's matrix. 

Here, as a first illustration of the method, we shall consider the large-$U$ regime of the Hubbard model where the construction of such domains is rather natural. 
Indeed, at large $U$ and half-filling, the Hubbard model approaches the Heisenberg limit where only the $2^N$ states with no double occupancy, $n_D(n)=0$, have a significant weight in the wave function.
The contribution of the other states vanishes as $U$ increases with a rate increasing sharply with $n_D(n)$. 
In addition, for a given number of double occupations, configurations with large values of $n_A(n)$ are favored due to their high electronic mobility.
Therefore, we build domains associated with small $n_D$ and large $n_A$ in a hierarchical way as described below. 

For simplicity and reducing the number of matrix inversions to perform, we shall consider only one non-trivial domain called here the main domain and denoted as $\cD$. 
This domain will be chosen common to all states belonging to it, that is,
\be
	\cD_i= \cD \qq{for} \ket{i} \in \cD.
\ee
For the other states, we choose a single-state domain as
\be
	\cD_i= \qty{ \ket{i} } \qq{for} \ket{i} \notin \cD.
\ee
To define $\cD$, let us introduce the following set of states
\be
	{\cS_{ij} = \qty{ \ket{n} : n_D(n)=i \land n_A(n)=j }}.
\ee
$\cD$ is defined as containing the set of states having up to $n_D^\text{max}$ double occupations and, for each state with a number of double occupations equal to $m$, a number of nearest-neighbor antiparallel pairs between $n_A^\text{min}(m)$ and $n_A^\text{max}(m)$. 
Here, $n_A^\text{max}(m)$ is fixed and set at its maximum value for a given $m$, \ie, $n_A^\text{max}(m)= \max(N-1-2m,0)$. 
Using these definitions, the main domain is taken as the union of some elementary domains
\be
	\cD = \bigcup_{n_D=0}^{n_D^\text{max}}\cD(n_D,n_A^\text{min}(n_D)),
\ee
where the elementary domains are defined as
\be
	\cD(n_D,n_A^\text{min}(n_D))=\bigcup_{ j = n_A^\text{min}(n_D)}^{n_A^\text{max}(n_D)} \cS_{n_D j}.
\ee
The two quantities defining the main domain are thus $n_D^\text{max}$ and $n_A^\text{min}(m)$.
To give an illustrative example, let us consider the 4-site case. 
There are 6 possible elementary domains:
\begin{align*}
	\cD(0,3) & = \cS_{03}, 
	&
	\cD(0,2) & = \cS_{03} \cup \cS_{02}, 
	\\
	\cD(0,1) & = \cS_{03} \cup \cS_{02} \cup \cS_{01},
	&
	\cD(1,1) & = \cS_{11},
	\\
	\cD(1,0) & = \cS_{11} \cup \cS_{10},
	&
	\cD(2,0) & = \cS_{20}, 
\end{align*}
where
\begin{align*}
	\cS_{03} & = \qty{ \ket{\uparrow,\downarrow,\uparrow,\downarrow }, \ket{\downarrow,\uparrow,\downarrow,\uparrow } }, 
	\qq{(the two N\'eel states)}
\\
	\cS_{02} & = \qty{ \ket{\uparrow, \downarrow, \downarrow, \uparrow }, \ket{\downarrow, \uparrow, \uparrow, \downarrow } },
\\
	\cS_{01} & = \qty{ \ket{\uparrow, \uparrow, \downarrow, \downarrow }, \ket{\downarrow, \downarrow, \uparrow, \uparrow } },
\\
	\cS_{11} & = \qty{ \ket{\uparrow \downarrow, \uparrow ,\downarrow, 0 }, \ket{\uparrow \downarrow, 0, \uparrow,\uparrow } + \ldots },
\\
	\cS_{10} & = \qty{ \ket{\uparrow \downarrow, \uparrow, 0, \downarrow }, \ket{\uparrow \downarrow, 0, \uparrow, \downarrow } + \ldots },
\\
	\cS_{20} & = \qty{ \ket{\uparrow \downarrow, \uparrow \downarrow, 0 ,0 } + \ldots }.
\end{align*}
For the three last cases, the dots indicate that one must also consider the remaining states obtained by permuting the position of the pairs.

\subsection{DMC simulations}

Let us now present our DMC calculations for the Hubbard model. 
In what follows, we shall restrict ourselves to the case of the Green's function Monte Carlo approach where trapping times are integrated out exactly.

Let us begin with a small chain of 4 sites with $U=12$. 
From now on, we shall take $t=1$. 
The size of the linear space is ${\binom{4}{2}}^2 = 36$ and the ground-state energy obtained by exact diagonalization is $E_0=-0.768068\ldots$. 
The two variational parameters of the trial vector have been optimized and fixed at the values of $\alpha=1.292$ and $\beta=0.552$ with a 
variational energy $E_\text{v}=-0.495361\ldots$. 
In what follows, $\ket{I_0}$ is systematically chosen as one of the two N\'eel states, \eg, $\ket{I_0} = \ket{\uparrow,\downarrow, \uparrow,\ldots}$. 

\begin{figure}
\includegraphics[width=\columnwidth]{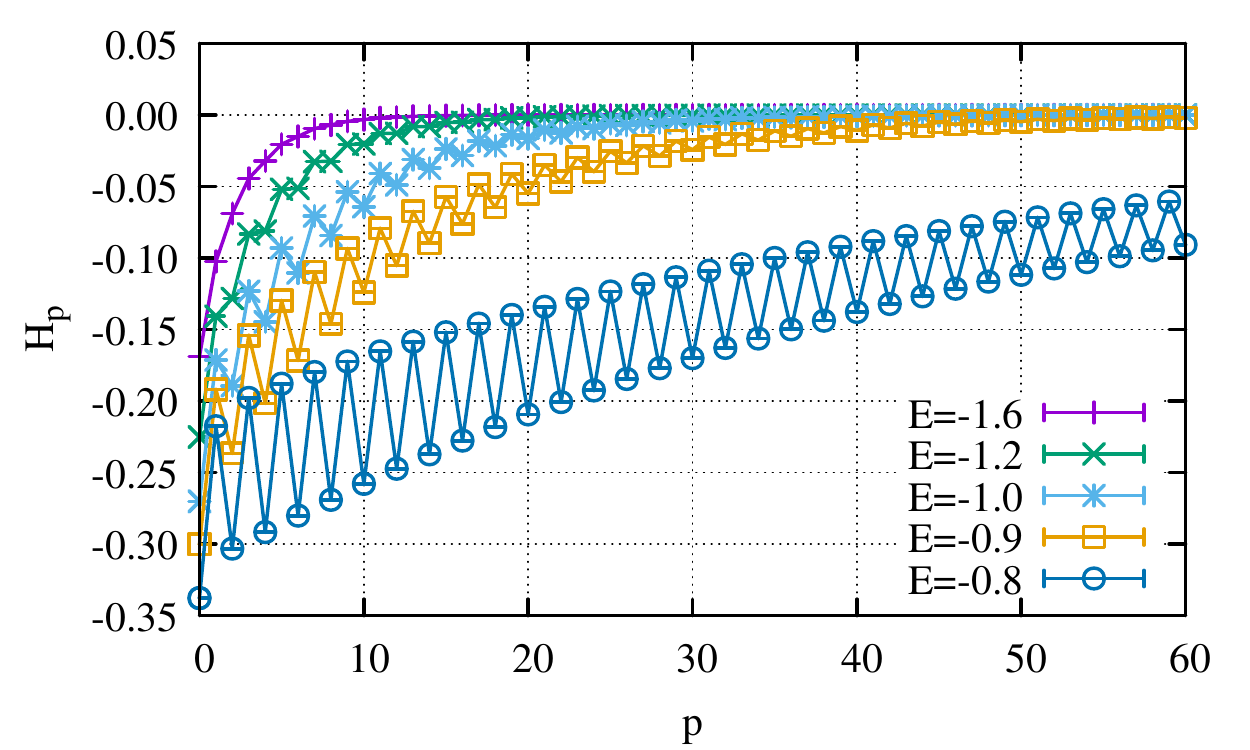}
\caption{One-dimensional Hubbard model with $N=4$ and $U=12$. 
$H_p$ as a function of $p$ for $E=-1.6$, $-1.2$, $-1.0$, $-0.9$, and $-0.8$. 
$H_0$ is computed analytically and $H_p$ ($p > 0$) is computed stochastically. 
Error bars are smaller than the symbol size.}
\label{fig3}
\end{figure}

Figure \ref{fig3} shows the convergence of $H^\text{DMC}_p$, Eq.~\eqref{eq:defHp}, as a function of $p$ for different values of the reference energy $E$.
We consider the simplest case where a single-state domain is associated to each state.
Five different values of $E$ have been chosen, namely $E=-1.6$, $-1.2$, $-1.0$, $-0.9$, and $-0.8$. 
Only $H_0$ is computed analytically ($p_\text{ex}=0$). 
At the scale of the figure, the error bars are too small to be seen.

\begin{figure}
\includegraphics[width=\columnwidth]{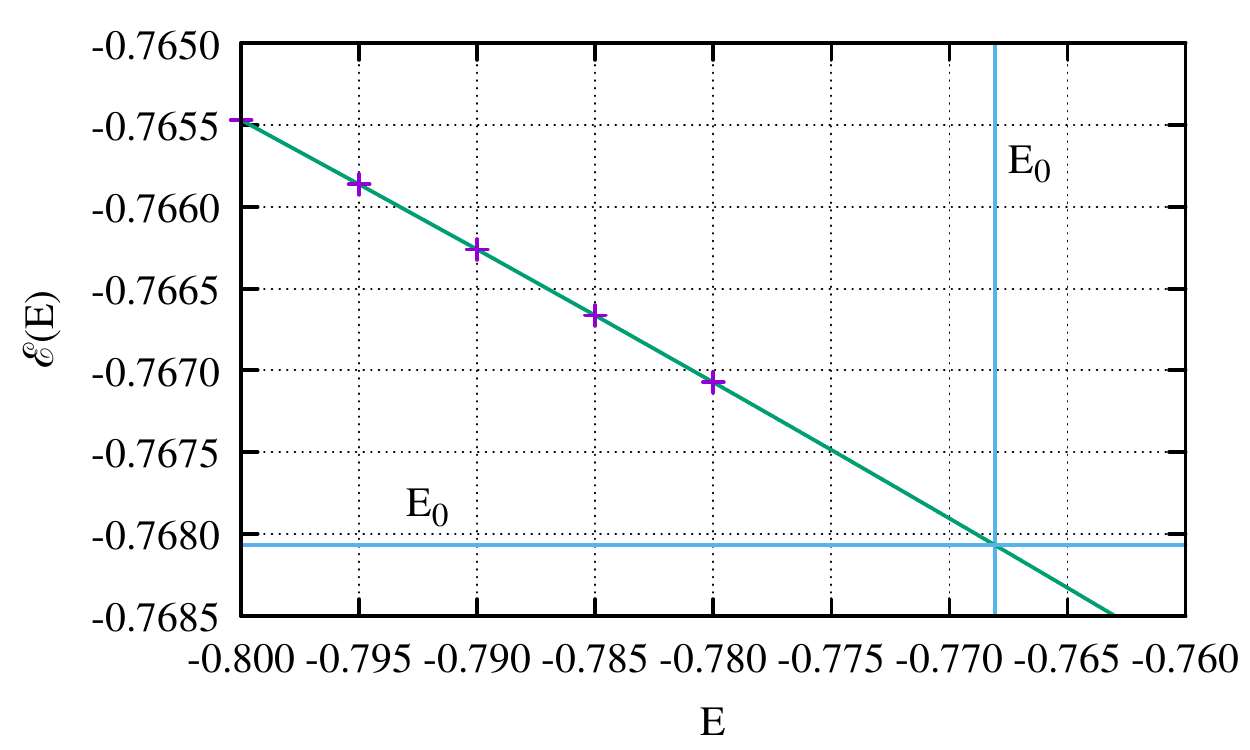}
\caption{One-dimensional Hubbard model with $N=4$ and $U=12$. 
$\cE(E)$ as a function of $E$.
The horizontal and vertical lines are at $\cE(E_0)=E_0$ and $E=E_0$, respectively.
$E_0 = -0.768068\ldots$ is the exact energy. 
The dotted line is the two-component extrapolation defined in Eq.~\eqref{eq:2comp}.
Error bars are smaller than the symbol size.}
\label{fig4}
\end{figure}

When $E$ is far from the exact value, the convergence is very rapid and only a few terms of the $p$-expansion are necessary. 
In contrast, when $E$ approaches the exact energy, a slower convergence is observed, as expected from the divergence of the matrix elements of the Green's matrix at $E=E_0$ where the expansion does not converge at all. 
The oscillations of the curves as a function of $p$ are due to a parity effect specific to this system. 
In practice, it is not too much of a problem since a smoothly convergent behavior is nevertheless observed for the even- and odd-parity curves.
The ratio, $\cE^\text{DMC}(E,p_\text{ex}=1,p_\text{max})$, Eq.~\eqref{eq:E_pex}, as a function of $E$ is presented in Fig.~\ref{fig4}. 
Here, $p_\text{max}$ is taken sufficiently large so that the convergence at large $p$ is reached. 
The values of $E$ are $-0.780$, $-0.790$, $-0.785$, $-0.780$, and $-0.775$. 
For small $E$, the curve is extrapolated using the so-called two-component expression:
\be
\label{eq:2comp}
	\cE(E) = \frac{ \frac{\epsilon_0 c_0}{\epsilon_0-E} + \frac{\epsilon_1 c_1}{\epsilon_1-E}}{\frac{c_0}{\epsilon_0-E} + \frac{c_1}{\epsilon_1-E} },
\ee
which considers only the first two terms of the exact expression of $\cE(E)$ [see Eq.~\eqref{eq:calE}].
Here, the fitting parameters that need to be determined are $c_0$, $\epsilon_0$, $c_1$, and $\epsilon_1$.
The estimate of the energy obtained from $\cE(E)=E$ is $-0.76807(5)$ in full agreement with the exact value of $-0.768068\ldots$.

Table \ref{tab1} illustrates the dependence of the Monte Carlo results upon the choice of the domain. 
The reference energy is $E=-1$.
The first column indicates the various domains consisting of the union of some elementary domains as explained above. 
The first line of the table gives the results when one uses a single-state domain for all states and the last one for the maximal domain containing the full linear space. 
The size of the various domains is given in the second column, the average trapping time for the state $\ket{I_0}$ in the third column, and an estimate of the speed of convergence of the $p$-expansion for the energy in the fourth column. 
To quantify the rate of convergence, we report the quantity, $p_\text{conv}$, defined as the smallest value of $p$ for which the energy is roughly stabilized 
to five decimal places. 
The smaller $p_\text{conv}$, the better the convergence is.
Although this is a rough estimate, it is sufficient here for our purpose.
As clearly seen, the speed of convergence is directly related to the magnitude of $\bar{t}_{I_0}$. 
The longer the stochastic trajectories remain trapped within the domain, the better the convergence. 
Of course, when the domain is chosen to be the full space, the average trapping time becomes infinite. 
Let us emphasize that the rate of convergence has no reason to be related to the size of the domain.
For example, the domain $\cD(0,3) \cup \cD(1,0)$ has a trapping time for the N\'eel state of $6.2$, while the domain $\cD(0,3) \cup \cD(1,1)$, having almost the same number of states (28 states), has an average trapping time about 6 times longer. 
Finally, the last column gives the energy obtained for $E=-1$. 
The energy is expected to be independent of the domain and to converge to a common value, which is indeed the case here. 
The exact value, $\cE(E=-1)=-0.75272390\ldots$, can be found in the last row of Table \ref{tab1} for the case of a domain corresponding to the full space.
In sharp contrast, the statistical error depends strongly on the type of domains used. 
As expected, the largest error of $3 \times 10^{-5}$ is obtained in the case of a single-state domain for all states. 
The smallest statistical error is obtained for the ``best'' domain having the largest average trapping time. 
Using this domain leads to a reduction in the statistical error as large as about three orders of magnitude, which nicely illustrates the critical importance of the domains employed.

\begin{table}
\centering
\caption{One-dimensional Hubbard model with $N=4$, $U=12$, $E=-1$, $\alpha=1.292$, $\beta=0.552$, and $p_\text{ex}=4$. 
The simulation is performed with 20 independent blocks and $10^5$ stochastic paths starting from the N\'eel state. 
$\bar{t}_{I_0}$ is the average trapping time for the N\'eel state. 
$p_\text{conv}$ is a measure of the convergence of $\cE^\text{DMC}$ as a function of $p$ 
(smallest value of $p$ for which the energy is roughly stabilized to five decimal places)
See text for more details.}
\label{tab1}
\begin{ruledtabular}
\begin{tabular}{lrrrl}
Domain & Size & $\bar{t}_{I_0}$ & $p_\text{conv}$ & \mcc{$\cE^\text{DMC}(E=-1)$} \\
\hline
Single & 1 & 0.026 & 88 &$-0.75276(3)$\\
$\cD(0,3)$ & 2 & 2.1  & 110 &$-0.75276(3)$\\
$\cD(0,2)$ & 4 & 2.1 & 106  &$-0.75275(2)$\\
$\cD(0,1)$ & 6 & 2.1&  82   &$-0.75274(3)$\\
$\cD(0,3)\cup\cD(1,1)$ &14 &4.0& 60 & $-0.75270(2)$\\
$\cD(0,3)\cup\cD(1,0)$ &26 &6.2& 45 & $-0.752730(7)$ \\
$\cD(0,2)\cup\cD(1,1)$ &16 &10.1 & 36 &$-0.75269(1)$\\
$\cD(0,2)\cup\cD(1,0)$ &28 &34.7  & 14&$-0.7527240(6)$\\
$\cD(0,1)\cup\cD(1,1)$ &18 & 10.1 & 28 &$-0.75269(1)$\\
$\cD(0,1)\cup\cD(1,0)$ &30 & 108.7 & 11&$-0.75272400(5)$ \\
$\cD(0,3)\cup\cD(1,1)\cup\cD$(2,0) &20 & 4.1 & 47 &$-0.75271(2)$\\
$\cD(0,3)\cup\cD(1,0)\cup\cD$(2,0) &32 & 6.5 & 39 &$-0.752725(3)$\\
$\cD(0,2)\cup\cD(1,1)\cup\cD$(2,0) &22 & 10.8 & 30 &$-0.75270(1)$\\
$\cD(0,2)\cup\cD(1,0)\cup\cD$(2,0) &34 & 52.5 & 13&$-0.7527236(2)$\\
$\cD(0,1)\cup\cD(1,1)\cup\cD$(2,0) & 24 & 10.8 & 26&$-0.75270(1)$\\
$\cD(0,1)\cup\cD(1,0)\cup\cD$(2,0) & 36 & $\infty$&1&$-0.75272390$\\
\end{tabular}
\end{ruledtabular}
\end{table}

As explained above, it is very advantageous to calculate exactly as many $(H_p,S_p)$ as possible in order to avoid statistical error on the largest components. 
Ideally, this should be done up to the value of $p$ for which the calculation of these quantities whose cost increases exponentially is possible.
Table \ref{tab2} shows the results both for the case of a single-state domain and for the domain having the largest average trapping time, namely $\cD(0,1) \cup \cD(1,1)$ (see Table \ref{tab1}). 
Table \ref{tab2} reports the statistical fluctuations of the energy for the simulation of Table \ref{tab1}. 
Results show that it is indeed worth computing exactly as many components as possible. 
For the single-state domain, the statistical error is reduced by a factor of two when passing from no analytical computation, $p_\text{ex}=0$, to the case where eight components for $H_p$ and $S_p$ are computed exactly. 
For the best domain, the impact is much more important with a huge reduction of about three orders of magnitude in the statistical error. 

Table \ref{tab3} reports the energies convergence as a function of $p$ alongside their statistical error on the last digit for $E=
-0.8$, $-0.795$, $-0.79$, $-0.785$, and $-0.78$. 
The values are displayed in Fig.~\ref{fig4}. 
As seen, the behavior of $\cE$ as a function of $E$ is almost perfectly linear. 
The extrapolated values obtained from the five values of the energy with three different fitting functions are reported. 
Fitting the data using a simple linear function leads to an energy of $-0.7680282(5)$ (to be compared with the exact value of $-0.768068\ldots$). 
A small bias of about $4 \times 10^{-5}$ is observed. 
This bias vanishes within the statistical error when resorting to more flexible fitting functions, such as a quadratic function of $E$ or the
two-component representation given by Eq.~\eqref{eq:2comp}.
Our final value is in full agreement with the exact value with about six decimal places.

Table \ref{tab4} shows the evolution of the average trapping times and extrapolated energies as a function of $U$ when using $\cD(0,1) \cup \cD(1,0)$ as the main domain. 
We also report the variational and exact energies together with the values of the optimized parameters of the trial wave function. 
As $U$ increases the configurations with zero or one double occupation become more and more predominant and the average trapping time increases. 
For very large values of $U$ (say, $U \ge 12$) the increase of $\bar{t}_{I_0}$ becomes particularly steep.

Finally, in Table \ref{tab5}, we report the results obtained for larger systems at $U=12$ for a chain size ranging from $N=4$ (36 states) 
to $N=12$ ($\sim 10^6$ states). 
No careful construction of domains maximizing the average trapping time has been performed, we have merely chosen domains of reasonable size (no more than 2682) by taking not too large number of double occupations (only, $n_D=0,1$) and not too small number of nearest-neighbor antiparallel pairs. 
As seen, as the number of sites increases, the average trapping time for the chosen domains decreases. 
This point is, of course, undesirable since the efficiency of the approach may gradually deteriorate when considering large systems. 
A more elaborate way of constructing domains is clearly desirable in this case.
The exact energies extrapolated using the two-component function are also reported. 
Similarly to what has been done for $N=4$, the extrapolation is performed using about five values of the reference energy. 
The extrapolated DMC energies are in full agreement with the exact value within the error bar. 
However, an increase in statistical error is observed when the system size increases. 
To get lower error bars, more accurate trial wave functions may be considered, better domains, and also larger simulation times. 
Of course, it will also be particularly interesting to take advantage of the fully parallelizable character of the algorithm to get much lower error bars. 
All these aspects will be considered in a forthcoming work.

\begin{table}
\caption{One-dimensional Hubbard model with $N=4$, $U=12$, and $E=-1$. 
Dependence of the statistical error on the energy with the number of $p$-components $p_\text{ex}$ calculated analytically in the expression 
of the energy, $\cE^\text{DMC}(E,p_\text{ex},p_\text{max})$, Eq.~\eqref{eq:E_pex}.
The simulation is performed the same way as in Table \ref{tab1}. 
Results are presented when a single-state domain is used for all states and when $\cD(0,1) \cup \cD(1,0)$ is chosen as the main domain.}
\label{tab2}
\begin{ruledtabular}
\begin{tabular}{lcc}
$p_{ex}$ & single-state & $\cD(0,1) \cup \cD(1,0)$ \\
\hline
$0$ & $4.3 \times 10^{-5}$ &$ 347 \times 10^{-8}$ \\
$1$ & $4.0  \times10^{-5}$ &$ 377 \times 10^{-8}$\\
$2$ & $3.7 \times 10^{-5}$ &$ 43  \times 10^{-8}$\\
$3$ & $3.3 \times 10^{-5}$ &$ 46  \times 10^{-8}$\\
$4$ & $2.6 \times 10^{-5}$ &$ 5.6  \times 10^{-8}$\\
$5$ & $2.5  \times10^{-5}$ &$ 6.0  \times 10^{-8}$\\
$6$ & $2.3  \times10^{-5}$ &$ 0.7  \times 10^{-8}$\\
$7$ & $2.2 \times 10^{-5}$ &$ 0.6  \times 10^{-8}$\\
$8$ & $2.2  \times10^{-5}$ &$  0.05 \times 10^{-8}$\\
\end{tabular}
\end{ruledtabular}
\end{table}

\begin{table}
\caption{One-dimensional Hubbard model with $N=4$, $U=12$, $\alpha=1.292$, $\beta=0.552$, and $p_\text{ex}=4$. 
The main domain is $\cD(0,1) \cup \cD(1,0)$. 
The simulation is performed with 20 independent blocks and $10^6$ stochastic paths starting from the N\'eel state. 
The fits are performed with the five values of $E$ reported in this table.}
\label{tab3}
\begin{ruledtabular}
\begin{tabular}{lc}
$E$ & $\cE^\text{DMC}(E)$ \\
\hline
$-0.8$  &$-0.7654686(2)$\\
$-0.795$&$-0.7658622(2)$\\
$-0.79$ &$-0.7662607(3)$\\
$-0.785$&$-0.7666642(4)$\\
$-0.78$ &$-0.7670729(5)$\\
$E_0$ linear fit &  $-0.7680282(5)$\\
$E_0$ quadratic fit & $-0.7680684(5)$\\
$E_0$ two-component fit & $-0.7680676(5)$\\
$E_0$ exact & $-0.768068\ldots$\\
\end{tabular}
\end{ruledtabular}
\end{table}

\begin{table}
\caption{One-dimensional Hubbard model with $N=4$. Average trapping time as a 
function of $U$. $(\alpha,\beta)$ and $E_\text{v}$ are the parameters and variational energy
of the trial wave function, respectively. $E_\text{ex}$ is the exact energy obtained by diagonalization of the 
Hamiltonian matrix.
The main domain is $\cD(0,1) \cup \cD(1,0)$.}
\label{tab4}
\begin{ruledtabular}
\begin{tabular}{rcccr}
$U$ & $(\alpha,\beta)$ & $E_\text{v}$ & $E_\text{ex}$  & $\bar{t}_{I_0}$ \\
\hline
8  & (0.908,\,0.520) & $-0.770342\ldots$ &$-1.117172\ldots$ & 33.5\\
10 & (1.116,\,0.539) & $-0.604162\ldots$ &$-0.911497\ldots$ & 63.3\\
12 & (1.292,\,0.552) & $-0.495361\ldots$ &$-0.768068\ldots$ & 108.7\\
14 & (1.438,\,0.563) & $-0.419163\ldots$ &$-0.662871\ldots$ & 171.7  \\
20 & (1.786,\,0.582) & $-0.286044\ldots$ &$-0.468619\ldots$ & 504.5  \\
50 & (2.690,\,0.609) & $-0.110013\ldots$ &$-0.188984\ldots$ & 8040.2  \\
200 & (4.070,\,0.624)& $-0.026940\ldots$ &$-0.047315\ldots$ & 523836.0  \\
\end{tabular}
\end{ruledtabular}
\end{table}

\begin{table*}
\caption{Ground-state energy of the one-dimensional Hubbard model for different sizes and $U=12$. 
The parameters $(\alpha,\beta)$ of the trial wave function and the corresponding variational energy $E_\text{v}$ are reported.
$\bar{t}_{I_0}$ is the average trapping time for the N\'eel state.
The extrapolated DMC energies, $E_\text{DMC}$, are obtained using the two-component fitting function, Eq.~\eqref{eq:2comp}. $E_\text{DMC}$ values are in full agreement with the exact values $E_\text{ex}$ computed by diagonalization of the Hamiltonian matrix.}
\label{tab5}
\begin{ruledtabular}
\begin{tabular}{crcrccccr}
$N$ & Hilbert space size & Domain & Domain size & $(\alpha,\beta)$ &$\bar{t}_{I_0}$ & $E_\text{v}$ & $E_\text{DMC}$ & $ E_\text{ex}$\\
\hline
4  & 36     & $\cD(0,1) \cup \cD(1,0)$ & 30 &(1.292,\,0.552)& 108.7  & $-0.495361$    &    $-0.7680676(5)$  &  $-0.768068$  \\
6  & 400    & $\cD(0,1) \cup \cD(1,0)$ &200 &(1.124,\,0.689)&57.8 & $-0.633297$       &     $-1.215389(9)$  &  $-1.215395$  \\
8  & 4 900   & $\cD(0,1) \cup \cD(1,0)$ & 1 190 &(0.984,\,0.788)& 42.8 & $-0.750995 $ &     $-1.6637(2)$    &  $-1.66395$  \\
10 & 63 504  & $\cD(0,5) \cup \cD(1,4)$ & 2 682 &(0.856,\,0.869)& 31.0  & $-0.855958$ &    $-2.1120(7)$     &  $-2.113089$ \\
12 & 853 776 & $\cD(0,8) \cup \cD(1,7)$ & 1 674 &(0.739,\,0.938)& 16.7 & $-0.952127$  &    $-2.560(6)$      &  $-2.562529$  \\
\end{tabular}
\end{ruledtabular}
\end{table*}

\section{Summary and perspectives}
\label{sec:conclu}

In this work, it has been shown how to integrate out exactly --- within a DMC framework --- the contribution of all stochastic trajectories trapped in some given domains of the Hilbert space. 
The corresponding equations have been derived in a general context.
In such a way, a new effective stochastic dynamics connecting only domains (and not the individual states) is defined.
A key property of this effective dynamics is that it does not depend on the ``shape'' of the domains used for each state.
Therefore, rather general domains (with or without overlap) can be considered. 

To obtain the effective transition probability (which provides the probability of going from one domain to another) 
and the corresponding renormalized estimators, the Green's functions restricted to each domain must be computed analytically, that is, in practice, matrices of the size of the number of states for the sampled domains have to be inverted. 
This is the main computationally intensive step of the present approach. 
The efficiency of the method is directly related to the importance of the average time spent by the stochastic trajectories in each domain. 

Being able to define domains with large average trapping times is the key aspect of the method since it may lead to some important reduction of the statistical error, as illustrated in our numerical applications.
Therefore, a trade-off has to be found between maximizing the average trapping time and minimizing the cost of computing the domain Green's functions. 
In practice, there is no general rule to construct such domains.  
For each system at hand, one needs to determine, on physical grounds, which regions of the configuration space are preferentially sampled by the stochastic trajectories and to build domains of minimal size enclosing such regions. 

In the first application presented here on the one-dimensional Hubbard model, we exploit the physics of the large-$U$ regime that is known to approach the Heisenberg limit where double occupations have small weights. 
This simple example has been chosen to illustrate the various aspects of the approach. 

Our goal is, of course, to tackle much larger systems, like those treated by state-of-the-art methods, such as selected CI, \cite{Huron_1973,Harrison_1991,Giner_2013,Holmes_2016,Schriber_2016,Tubman_2020}, FCIQMC, \cite{Booth_2009,Cleland_2010}, AFQMC, \cite{Zhang_2003} or DMRG. \cite{White_1999,Chan_2011}
Here, we have mainly focused on the theoretical aspects of the approach. 
In order to consider larger systems, an elaborate implementation of the present method is necessary in order to keep under control the cost of the simulation.
This is outside the scope of the present study and will be presented in a forthcoming work.

\acknowledgments{
P.F.L., A.S., and M.C. acknowledge funding from the European Research Council (ERC) under the European Union's Horizon 2020 research and innovation programme (Grant agreement No.~863481).
This work was supported by the European Centre of Excellence in Exascale Computing TREX --- Targeting Real Chemical Accuracy at the Exascale. 
This project has received funding from the European Union's Horizon 2020 --- Research and Innovation program --- under grant agreement no.~952165.
}

\appendix

\section{Particular case of a $2\times2$ matrix}
\label{app:A}

For the simplest case of a system containing only two states, $\ket{1}$ and $\ket{2}$, the fundamental equation given in Eq.~\eqref{eq:eqfond} simplifies to
\be
	\cI
	= \mel{ I_0 }{ \qty(H-E \Id)^{-1} }{ \Psi } 
	=  \mel{ I_0 }{ { \qty[ P_0 \qty(H-E \Id) P_0 ]}^{-1} }{ \Psi }
	+ \sum_{p=1}^{\infty} \cI_p,
\ee
with
\begin{multline}
	\cI_p= \sum_{I_1 \notin \cD_0, \ldots , I_p \notin \cD_{p-1}}
	\qty[ \prod_{k=0}^{p-1} \mel{ I_k }{ { \qty[P_k \qty(H-E \Id) P_k ]}^{-1} (-H)(1-P_k) }{ I_{k+1} } ]
	\\
	\times \mel{ I_p }{ { \qty[ P_p \qty(H-E \Id) P_p ]}^{-1} }{ \Psi }.
\end{multline}
To treat simultaneously the two possibilities for the final state, \ie, $\ket{i_N} = \ket{1}$ or $\ket{2}$, Eq.~\eqref{eq:eqfond} has been slightly generalized to the case of a general vector for the final state written as
\be
	\ket{\Psi} = \Psi_1 \ket{1} +  \Psi_2 \ket{2}.
\ee

Let us choose a single-state domain for both states, namely $\cD_1 = \qty{ \ket{1} }$  and $\cD_2 = \qty{ \ket{2} }$. 
Note that, due to the simplicity of the present two-state model, there are only two  possible deterministic ``alternating'' paths, namely, $\ket{1} \to \ket{2} \to  \ket{1},\ldots$ and $\ket{2} \to \ket{1} \to \ket{2},\ldots$.

For the sake of convenience, we introduce the following quantities
\begin{align}
\label{eq:defA1}
	A_1 & = \mel{ 1 }{ {\qty[ P_1 \qty(H-E \Id) P_1 ]}^{-1} (-H)(1-P_1) }{ 2 },
	\\
\label{eq:defA2}
	A_2 & = \mel{ 2 }{ {\qty[ P_2 \qty(H-E \Id) P_2]}^{-1} (-H) (1-P_2) }{ 1 }, 
\end{align}
and
\begin{align}
	C_1 & = \mel{ 1 }{ {\qty[ P_1 \qty(H-E \Id) P_1]}^{-1}  }{ \Psi },
	\\
	C_2 & = \mel{ 2 }{ {\qty[ P_2 \qty(H-E \Id) P_2]}^{-1} }{ \Psi }.
\end{align}
Without loss of generality, let us choose, for example, $\ket{I_0} = \ket{1}$.
Then, it is easy to show that
\begin{align}
	\cI_{2k+1} & = \frac{C_2}{A_2} (A_1 A_2)^{2k+1},
	& 
	\cI_{2k} & = C_1 (A_1 A_2)^{2k},
\end{align}
which yields
\be
	\sum_{p=1}^{\infty}
	\cI_p
	= \frac{C_2}{A_2} \sum_{p=1}^{\infty} (A_1 A_2)^p  + C_1  \sum_{p=1}^{\infty} (A_1 A_2)^p
	= A_1 \frac{C_2 + C_1 A_2}{1-A_1 A_2},
\ee
and 
\be
	\mel{ 1 }{ \qty(H-E \Id)^{-1} }{ \Psi } 
	= C_1 +  A_1 \frac{C_2 + C_1 A_2}{1-A_1 A_2}.
\ee
For a $2\times2$ matrix of the form
\be
\label{eq:2x2_matrix}
	H = 
	\begin{pmatrix}
	H_{11}	&	H_{12}	\\
	H_{12}	&	H_{22}	\\
	\end{pmatrix},
\ee
it is easy to evaluate the $A_i$'s.
Using Eqs.~\eqref{eq:defA1} and \eqref{eq:defA2}, one gets, for $i = 1$ or $2$,
\begin{align}
	A_i & = -\frac{H_{12}}{H_{ii}-E},
	&
	C_i & = \frac{1}{H_{ii}-E} \Psi_i,
\end{align}
which finally yields
\be
	\mel{ 1 }{ \qty(H-E \Id)^{-1} }{ \Psi} 
	= \frac{ H_{22}-E}{\Delta}  \Psi_1 - \frac{H_{12}}{\Delta} \Psi_2,
\label{eq:final}
\ee
where $\Delta$ is the determinant of $H$. 

Alternatively, the quantity $\mel{ 1 }{ \qty(H-E \Id)^{-1} }{ \Psi}$ in the left-hand-side of Eq.~\eqref{eq:final} can be directly calculated using the inverse of the matrix defined in Eq.~\eqref{eq:2x2_matrix}, yielding
\be
\begin{split}
\qty(H-E \Id)^{-1} \ket{\Psi}& =
              \frac{1}{H-E \Id}      \begin{pmatrix}
         \Psi_1 \\
         \Psi_2   \\
        \end{pmatrix}
        \\
 & = \frac{1}{\Delta}
        \begin{pmatrix}
        H_{22}-E  &      - H_{21}  \\
        - H_{21}  &       H_{11}-E  \\
        \end{pmatrix}
\begin{pmatrix}
         \Psi_1 \\
         \Psi_2   \\
        \end{pmatrix}.
\end{split}
\ee
As readily seen, the first component of the vector $\qty(H-E \Id)^{-1}\ket{\Psi}$ is identical to the one given in Eq.~\eqref{eq:final}, thus confirming independently the validity of this equation.

\bibliography{g}

\begin{thebibliography}{41}%
\makeatletter
\providecommand \@ifxundefined [1]{%
 \@ifx{#1\undefined}
}%
\providecommand \@ifnum [1]{%
 \ifnum #1\expandafter \@firstoftwo
 \else \expandafter \@secondoftwo
 \fi
}%
\providecommand \@ifx [1]{%
 \ifx #1\expandafter \@firstoftwo
 \else \expandafter \@secondoftwo
 \fi
}%
\providecommand \natexlab [1]{#1}%
\providecommand \enquote  [1]{``#1''}%
\providecommand \bibnamefont  [1]{#1}%
\providecommand \bibfnamefont [1]{#1}%
\providecommand \citenamefont [1]{#1}%
\providecommand \href@noop [0]{\@secondoftwo}%
\providecommand \href [0]{\begingroup \@sanitize@url \@href}%
\providecommand \@href[1]{\@@startlink{#1}\@@href}%
\providecommand \@@href[1]{\endgroup#1\@@endlink}%
\providecommand \@sanitize@url [0]{\catcode `\\12\catcode `\$12\catcode
  `\&12\catcode `\#12\catcode `\^12\catcode `\_12\catcode `\%12\relax}%
\providecommand \@@startlink[1]{}%
\providecommand \@@endlink[0]{}%
\providecommand \url  [0]{\begingroup\@sanitize@url \@url }%
\providecommand \@url [1]{\endgroup\@href {#1}{\urlprefix }}%
\providecommand \urlprefix  [0]{URL }%
\providecommand \Eprint [0]{\href }%
\providecommand \doibase [0]{http://dx.doi.org/}%
\providecommand \selectlanguage [0]{\@gobble}%
\providecommand \bibinfo  [0]{\@secondoftwo}%
\providecommand \bibfield  [0]{\@secondoftwo}%
\providecommand \translation [1]{[#1]}%
\providecommand \BibitemOpen [0]{}%
\providecommand \bibitemStop [0]{}%
\providecommand \bibitemNoStop [0]{.\EOS\space}%
\providecommand \EOS [0]{\spacefactor3000\relax}%
\providecommand \BibitemShut  [1]{\csname bibitem#1\endcsname}%
\let\auto@bib@innerbib\@empty
\bibitem [{\citenamefont {Foulkes}\ \emph {et~al.}(2001)\citenamefont
  {Foulkes}, \citenamefont {Mitas}, \citenamefont {Needs},\ and\ \citenamefont
  {Rajagopal}}]{Foulkes_2001}%
  \BibitemOpen
  \bibfield  {author} {\bibinfo {author} {\bibfnamefont {W.~M.~C.}\
  \bibnamefont {Foulkes}}, \bibinfo {author} {\bibfnamefont {L.}~\bibnamefont
  {Mitas}}, \bibinfo {author} {\bibfnamefont {R.~J.}\ \bibnamefont {Needs}}, \
  and\ \bibinfo {author} {\bibfnamefont {G.}~\bibnamefont {Rajagopal}},\ }\href
  {\doibase 10.1103/RevModPhys.73.33} {\bibfield  {journal} {\bibinfo
  {journal} {Rev. Mod. Phys.}\ }\textbf {\bibinfo {volume} {73}},\ \bibinfo
  {pages} {33} (\bibinfo {year} {2001})}\BibitemShut {NoStop}%
\bibitem [{\citenamefont {Kolorenc}\ and\ \citenamefont
  {Mitas}(2011)}]{Kolorenc_2011}%
  \BibitemOpen
  \bibfield  {author} {\bibinfo {author} {\bibfnamefont {J.}~\bibnamefont
  {Kolorenc}}\ and\ \bibinfo {author} {\bibfnamefont {L.}~\bibnamefont
  {Mitas}},\ }\href {\doibase 10.1088/0034-4885/74/2/026502} {\bibfield
  {journal} {\bibinfo  {journal} {Rep. Prog. Phys.}\ }\textbf {\bibinfo
  {volume} {74}},\ \bibinfo {pages} {026502} (\bibinfo {year}
  {2011})}\BibitemShut {NoStop}%
\bibitem [{\citenamefont {Holzmann}\ \emph {et~al.}(2006)\citenamefont
  {Holzmann}, \citenamefont {Bernu},\ and\ \citenamefont
  {Ceperley}}]{Holzmann_2006}%
  \BibitemOpen
  \bibfield  {author} {\bibinfo {author} {\bibfnamefont {M.}~\bibnamefont
  {Holzmann}}, \bibinfo {author} {\bibfnamefont {B.}~\bibnamefont {Bernu}}, \
  and\ \bibinfo {author} {\bibfnamefont {D.~M.}\ \bibnamefont {Ceperley}},\
  }\href {\doibase 10.1103/PhysRevB.74.104510} {\bibfield  {journal} {\bibinfo
  {journal} {Phys. Rev. B}\ }\textbf {\bibinfo {volume} {74}},\ \bibinfo
  {pages} {104510} (\bibinfo {year} {2006})}\BibitemShut {NoStop}%
\bibitem [{\citenamefont {Carlson}\ \emph {et~al.}(2015)\citenamefont
  {Carlson}, \citenamefont {Gandolfi}, \citenamefont {Pederiva}, \citenamefont
  {Pieper}, \citenamefont {Schiavilla}, \citenamefont {Schmidt},\ and\
  \citenamefont {Wiringa}}]{Carlson_2015}%
  \BibitemOpen
  \bibfield  {author} {\bibinfo {author} {\bibfnamefont {J.}~\bibnamefont
  {Carlson}}, \bibinfo {author} {\bibfnamefont {S.}~\bibnamefont {Gandolfi}},
  \bibinfo {author} {\bibfnamefont {F.}~\bibnamefont {Pederiva}}, \bibinfo
  {author} {\bibfnamefont {S.~C.}\ \bibnamefont {Pieper}}, \bibinfo {author}
  {\bibfnamefont {R.}~\bibnamefont {Schiavilla}}, \bibinfo {author}
  {\bibfnamefont {K.~E.}\ \bibnamefont {Schmidt}}, \ and\ \bibinfo {author}
  {\bibfnamefont {R.~B.}\ \bibnamefont {Wiringa}},\ }\href {\doibase
  10.1103/RevModPhys.87.1067} {\bibfield  {journal} {\bibinfo  {journal} {Rev.
  Mod. Phys.}\ }\textbf {\bibinfo {volume} {87}},\ \bibinfo {pages} {1067}
  (\bibinfo {year} {2015})}\BibitemShut {NoStop}%
\bibitem [{\citenamefont {Carlson}(2007)}]{Carlson_2007}%
  \BibitemOpen
  \bibfield  {author} {\bibinfo {author} {\bibfnamefont {J.}~\bibnamefont
  {Carlson}},\ }\href {\doibase 10.1016/j.nuclphysa.2006.12.079} {\bibfield
  {journal} {\bibinfo  {journal} {Nucl. Physics. A}\ }\textbf {\bibinfo
  {volume} {787}},\ \bibinfo {pages} {516} (\bibinfo {year}
  {2007})}\BibitemShut {NoStop}%
\bibitem [{\citenamefont {Austin}\ \emph {et~al.}(2012)\citenamefont {Austin},
  \citenamefont {Zubarev},\ and\ \citenamefont {Lester}}]{Austin_2012}%
  \BibitemOpen
  \bibfield  {author} {\bibinfo {author} {\bibfnamefont {B.~M.}\ \bibnamefont
  {Austin}}, \bibinfo {author} {\bibfnamefont {D.~Y.}\ \bibnamefont {Zubarev}},
  \ and\ \bibinfo {author} {\bibfnamefont {W.~A.}\ \bibnamefont {Lester}},\
  }\href {\doibase 10.1021/cr2001564} {\bibfield  {journal} {\bibinfo
  {journal} {Chem. Rev.}\ }\textbf {\bibinfo {volume} {112}},\ \bibinfo {pages}
  {263} (\bibinfo {year} {2012})}\BibitemShut {NoStop}%
\bibitem [{\citenamefont {Golub}\ and\ \citenamefont
  {Van~Loan}(2012)}]{Golub_2012}%
  \BibitemOpen
  \bibfield  {author} {\bibinfo {author} {\bibfnamefont {G.~H.}\ \bibnamefont
  {Golub}}\ and\ \bibinfo {author} {\bibfnamefont {C.~F.}\ \bibnamefont
  {Van~Loan}},\ }\href@noop {} {\emph {\bibinfo {title} {Matrix
  Computations}}},\ \bibinfo {edition} {4th}\ ed.\ (\bibinfo  {publisher} {The
  Johns Hopkins University Press},\ \bibinfo {year} {2012})\BibitemShut
  {NoStop}%
\bibitem [{\citenamefont {Davidson}(1975)}]{Davidson_1975}%
  \BibitemOpen
  \bibfield  {author} {\bibinfo {author} {\bibfnamefont {E.~R.}\ \bibnamefont
  {Davidson}},\ }\href {\doibase 10.1016/0021-9991(75)90065-0} {\bibfield
  {journal} {\bibinfo  {journal} {J. Comput. Phys.}\ }\textbf {\bibinfo
  {volume} {17}},\ \bibinfo {pages} {87} (\bibinfo {year} {1975})}\BibitemShut
  {NoStop}%
\bibitem [{\citenamefont {Caffarel}\ and\ \citenamefont
  {Claverie}(1988)}]{Caffarel_1988}%
  \BibitemOpen
  \bibfield  {author} {\bibinfo {author} {\bibfnamefont {M.}~\bibnamefont
  {Caffarel}}\ and\ \bibinfo {author} {\bibfnamefont {P.}~\bibnamefont
  {Claverie}},\ }\href {\doibase 10.1063/1.454227} {\bibfield  {journal}
  {\bibinfo  {journal} {J. Chem. Phys.}\ }\textbf {\bibinfo {volume} {88}},\
  \bibinfo {pages} {1088} (\bibinfo {year} {1988})}\BibitemShut {NoStop}%
\bibitem [{\citenamefont {Reynolds}\ \emph {et~al.}(1982)\citenamefont
  {Reynolds}, \citenamefont {Ceperley}, \citenamefont {Alder},\ and\
  \citenamefont {Lester}}]{Reynolds_1982}%
  \BibitemOpen
  \bibfield  {author} {\bibinfo {author} {\bibfnamefont {P.~J.}\ \bibnamefont
  {Reynolds}}, \bibinfo {author} {\bibfnamefont {D.~M.}\ \bibnamefont
  {Ceperley}}, \bibinfo {author} {\bibfnamefont {B.~J.}\ \bibnamefont {Alder}},
  \ and\ \bibinfo {author} {\bibfnamefont {W.~A.}\ \bibnamefont {Lester}},\
  }\href {\doibase 10.1063/1.443766} {\bibfield  {journal} {\bibinfo  {journal}
  {J. Chem. Phys.}\ }\textbf {\bibinfo {volume} {77}},\ \bibinfo {pages} {5593}
  (\bibinfo {year} {1982})}\BibitemShut {NoStop}%
\bibitem [{\citenamefont {Baroni}\ and\ \citenamefont
  {Moroni}(1999)}]{Baroni_1999}%
  \BibitemOpen
  \bibfield  {author} {\bibinfo {author} {\bibfnamefont {S.}~\bibnamefont
  {Baroni}}\ and\ \bibinfo {author} {\bibfnamefont {S.}~\bibnamefont
  {Moroni}},\ }\href {\doibase 10.1103/PhysRevLett.82.4745} {\bibfield
  {journal} {\bibinfo  {journal} {Phys. Rev. Lett.}\ }\textbf {\bibinfo
  {volume} {82}},\ \bibinfo {pages} {4745} (\bibinfo {year}
  {1999})}\BibitemShut {NoStop}%
\bibitem [{\citenamefont {Sorella}(1998)}]{Sorella_1998}%
  \BibitemOpen
  \bibfield  {author} {\bibinfo {author} {\bibfnamefont {S.}~\bibnamefont
  {Sorella}},\ }\href {\doibase 10.1103/PhysRevLett.80.4558} {\bibfield
  {journal} {\bibinfo  {journal} {Phys. Rev. Lett.}\ }\textbf {\bibinfo
  {volume} {80}},\ \bibinfo {pages} {4558} (\bibinfo {year}
  {1998})}\BibitemShut {NoStop}%
\bibitem [{\citenamefont {Assaraf}\ \emph {et~al.}(2000)\citenamefont
  {Assaraf}, \citenamefont {Caffarel},\ and\ \citenamefont
  {Khelif}}]{Assaraf_2000}%
  \BibitemOpen
  \bibfield  {author} {\bibinfo {author} {\bibfnamefont {R.}~\bibnamefont
  {Assaraf}}, \bibinfo {author} {\bibfnamefont {M.}~\bibnamefont {Caffarel}}, \
  and\ \bibinfo {author} {\bibfnamefont {A.}~\bibnamefont {Khelif}},\ }\href
  {\doibase 10.1103/physreve.61.4566} {\bibfield  {journal} {\bibinfo
  {journal} {Phys. Rev. E}\ }\textbf {\bibinfo {volume} {61}},\ \bibinfo
  {pages} {4566} (\bibinfo {year} {2000})}\BibitemShut {NoStop}%
\bibitem [{\citenamefont {Assaraf}\ and\ \citenamefont
  {Caffarel}(1999)}]{Assaraf_1999A}%
  \BibitemOpen
  \bibfield  {author} {\bibinfo {author} {\bibfnamefont {R.}~\bibnamefont
  {Assaraf}}\ and\ \bibinfo {author} {\bibfnamefont {M.}~\bibnamefont
  {Caffarel}},\ }\href {\doibase 10.1103/PhysRevLett.83.4682} {\bibfield
  {journal} {\bibinfo  {journal} {Phys. Rev. Lett.}\ }\textbf {\bibinfo
  {volume} {83}},\ \bibinfo {pages} {4682} (\bibinfo {year}
  {1999})}\BibitemShut {NoStop}%
\bibitem [{\citenamefont {Assaraf}\ \emph {et~al.}(1999)\citenamefont
  {Assaraf}, \citenamefont {Azaria}, \citenamefont {Caffarel},\ and\
  \citenamefont {Lecheminant}}]{Assaraf_1999B}%
  \BibitemOpen
  \bibfield  {author} {\bibinfo {author} {\bibfnamefont {R.}~\bibnamefont
  {Assaraf}}, \bibinfo {author} {\bibfnamefont {P.}~\bibnamefont {Azaria}},
  \bibinfo {author} {\bibfnamefont {M.}~\bibnamefont {Caffarel}}, \ and\
  \bibinfo {author} {\bibfnamefont {P.}~\bibnamefont {Lecheminant}},\ }\href
  {\doibase 10.1103/PhysRevB.60.2299} {\bibfield  {journal} {\bibinfo
  {journal} {Phys. Rev. B}\ }\textbf {\bibinfo {volume} {60}},\ \bibinfo
  {pages} {2299} (\bibinfo {year} {1999})}\BibitemShut {NoStop}%
\bibitem [{\citenamefont {Caffarel}\ and\ \citenamefont
  {Assaraf}(2000)}]{Caffarel_2000}%
  \BibitemOpen
  \bibfield  {author} {\bibinfo {author} {\bibfnamefont {M.}~\bibnamefont
  {Caffarel}}\ and\ \bibinfo {author} {\bibfnamefont {R.}~\bibnamefont
  {Assaraf}},\ }in\ \href@noop {} {\emph {\bibinfo {booktitle} {Lecture Notes
  in Chemistry}}},\ \bibinfo {editor} {edited by\ \bibinfo {editor}
  {\bibfnamefont {M.}~\bibnamefont {Defranceschi}}\ and\ \bibinfo {editor}
  {\bibfnamefont {C.~L.}\ \bibnamefont {Bris}}}\ (\bibinfo  {publisher}
  {Springer},\ \bibinfo {year} {2000})\ p.~\bibinfo {pages} {45}\BibitemShut
  {NoStop}%
\bibitem [{\citenamefont {Kalos}(1962)}]{Kalos_1962}%
  \BibitemOpen
  \bibfield  {author} {\bibinfo {author} {\bibfnamefont {M.~H.}\ \bibnamefont
  {Kalos}},\ }\href {\doibase 10.1103/PhysRev.128.1791} {\bibfield  {journal}
  {\bibinfo  {journal} {Phys. Rev.}\ }\textbf {\bibinfo {volume} {128}},\
  \bibinfo {pages} {1791} (\bibinfo {year} {1962})}\BibitemShut {NoStop}%
\bibitem [{\citenamefont {Kalos}(1970)}]{Kalos_1970}%
  \BibitemOpen
  \bibfield  {author} {\bibinfo {author} {\bibfnamefont {M.~H.}\ \bibnamefont
  {Kalos}},\ }\href {\doibase 10.1103/PhysRevA.2.250} {\bibfield  {journal}
  {\bibinfo  {journal} {Phys. Rev. A}\ }\textbf {\bibinfo {volume} {2}},\
  \bibinfo {pages} {250} (\bibinfo {year} {1970})}\BibitemShut {NoStop}%
\bibitem [{\citenamefont {Kalos}\ \emph {et~al.}(1974)\citenamefont {Kalos},
  \citenamefont {Levesque},\ and\ \citenamefont {Verlet}}]{Kalos_1974}%
  \BibitemOpen
  \bibfield  {author} {\bibinfo {author} {\bibfnamefont {M.~H.}\ \bibnamefont
  {Kalos}}, \bibinfo {author} {\bibfnamefont {D.}~\bibnamefont {Levesque}}, \
  and\ \bibinfo {author} {\bibfnamefont {L.}~\bibnamefont {Verlet}},\ }\href
  {\doibase 10.1103/PhysRevA.9.2178} {\bibfield  {journal} {\bibinfo  {journal}
  {Phys. Rev. A}\ }\textbf {\bibinfo {volume} {9}},\ \bibinfo {pages} {2178}
  (\bibinfo {year} {1974})}\BibitemShut {NoStop}%
\bibitem [{\citenamefont {Ceperley}\ and\ \citenamefont
  {Kalos}(1979)}]{Ceperley_1979}%
  \BibitemOpen
  \bibfield  {author} {\bibinfo {author} {\bibfnamefont {D.}~\bibnamefont
  {Ceperley}}\ and\ \bibinfo {author} {\bibfnamefont {M.}~\bibnamefont
  {Kalos}},\ }\href@noop {} {\emph {\bibinfo {title} {Monte Carlo Methods in
  Statistical Physics}}},\ edited by\ \bibinfo {editor} {\bibnamefont
  {K.Binder}}\ (\bibinfo  {publisher} {Springer, Berlin},\ \bibinfo {year}
  {1979})\ Chap.~\bibinfo {chapter} {4}\BibitemShut {NoStop}%
\bibitem [{\citenamefont {Ceperley}(1983)}]{Ceperley_1983}%
  \BibitemOpen
  \bibfield  {author} {\bibinfo {author} {\bibfnamefont {D.}~\bibnamefont
  {Ceperley}},\ }\href {\doibase https://doi.org/10.1016/0021-9991(83)90161-4}
  {\bibfield  {journal} {\bibinfo  {journal} {J. Comput. Phys.}\ }\textbf
  {\bibinfo {volume} {51}},\ \bibinfo {pages} {404} (\bibinfo {year}
  {1983})}\BibitemShut {NoStop}%
\bibitem [{\citenamefont {Moskowitz}\ and\ \citenamefont
  {Schmidt}(1986)}]{Moskowitz_1986}%
  \BibitemOpen
  \bibfield  {author} {\bibinfo {author} {\bibfnamefont {J.~W.}\ \bibnamefont
  {Moskowitz}}\ and\ \bibinfo {author} {\bibfnamefont {K.~E.}\ \bibnamefont
  {Schmidt}},\ }\href {\doibase 10.1063/1.451046} {\bibfield  {journal}
  {\bibinfo  {journal} {J. Chem. Phys.}\ }\textbf {\bibinfo {volume} {85}},\
  \bibinfo {pages} {2868} (\bibinfo {year} {1986})}\BibitemShut {NoStop}%
\bibitem [{\citenamefont {Petruzielo}\ \emph {et~al.}(2012)\citenamefont
  {Petruzielo}, \citenamefont {Holmes}, \citenamefont {Changlani},
  \citenamefont {Nightingale},\ and\ \citenamefont
  {Umrigar}}]{Petruzielo_2012}%
  \BibitemOpen
  \bibfield  {author} {\bibinfo {author} {\bibfnamefont {F.~R.}\ \bibnamefont
  {Petruzielo}}, \bibinfo {author} {\bibfnamefont {A.~A.}\ \bibnamefont
  {Holmes}}, \bibinfo {author} {\bibfnamefont {H.~J.}\ \bibnamefont
  {Changlani}}, \bibinfo {author} {\bibfnamefont {M.~P.}\ \bibnamefont
  {Nightingale}}, \ and\ \bibinfo {author} {\bibfnamefont {C.~J.}\ \bibnamefont
  {Umrigar}},\ }\href {\doibase 10.1103/PhysRevLett.109.230201} {\bibfield
  {journal} {\bibinfo  {journal} {Phys. Rev. Lett.}\ }\textbf {\bibinfo
  {volume} {109}},\ \bibinfo {pages} {230201} (\bibinfo {year}
  {2012})}\BibitemShut {NoStop}%
\bibitem [{\citenamefont {Garniron}\ \emph {et~al.}(2017)\citenamefont
  {Garniron}, \citenamefont {Scemama}, \citenamefont {Loos},\ and\
  \citenamefont {Caffarel}}]{Garniron_2017b}%
  \BibitemOpen
  \bibfield  {author} {\bibinfo {author} {\bibfnamefont {Y.}~\bibnamefont
  {Garniron}}, \bibinfo {author} {\bibfnamefont {A.}~\bibnamefont {Scemama}},
  \bibinfo {author} {\bibfnamefont {P.-F.}\ \bibnamefont {Loos}}, \ and\
  \bibinfo {author} {\bibfnamefont {M.}~\bibnamefont {Caffarel}},\ }\href
  {\doibase 10.1063/1.4992127} {\bibfield  {journal} {\bibinfo  {journal} {J.
  Chem. Phys.}\ }\textbf {\bibinfo {volume} {147}},\ \bibinfo {pages} {034101}
  (\bibinfo {year} {2017})}\BibitemShut {NoStop}%
\bibitem [{\citenamefont {Sharma}\ \emph {et~al.}(2017)\citenamefont {Sharma},
  \citenamefont {Holmes}, \citenamefont {Jeanmairet}, \citenamefont {Alavi},\
  and\ \citenamefont {Umrigar}}]{Sharma_2017}%
  \BibitemOpen
  \bibfield  {author} {\bibinfo {author} {\bibfnamefont {S.}~\bibnamefont
  {Sharma}}, \bibinfo {author} {\bibfnamefont {A.~A.}\ \bibnamefont {Holmes}},
  \bibinfo {author} {\bibfnamefont {G.}~\bibnamefont {Jeanmairet}}, \bibinfo
  {author} {\bibfnamefont {A.}~\bibnamefont {Alavi}}, \ and\ \bibinfo {author}
  {\bibfnamefont {C.~J.}\ \bibnamefont {Umrigar}},\ }\href {\doibase
  10.1021/acs.jctc.6b01028} {\bibfield  {journal} {\bibinfo  {journal} {J.
  Chem. Theory Comput.}\ }\textbf {\bibinfo {volume} {13}},\ \bibinfo {pages}
  {1595} (\bibinfo {year} {2017})}\BibitemShut {NoStop}%
\bibitem [{\citenamefont {Willow}\ \emph {et~al.}(2012)\citenamefont {Willow},
  \citenamefont {Kim},\ and\ \citenamefont {Hirata}}]{Willow_2012}%
  \BibitemOpen
  \bibfield  {author} {\bibinfo {author} {\bibfnamefont {S.~Y.}\ \bibnamefont
  {Willow}}, \bibinfo {author} {\bibfnamefont {K.~S.}\ \bibnamefont {Kim}}, \
  and\ \bibinfo {author} {\bibfnamefont {S.}~\bibnamefont {Hirata}},\ }\href
  {\doibase 10.1063/1.4768697} {\bibfield  {journal} {\bibinfo  {journal} {J.
  Chem. Phys.}\ }\textbf {\bibinfo {volume} {137}},\ \bibinfo {pages} {204122}
  (\bibinfo {year} {2012})}\BibitemShut {NoStop}%
\bibitem [{\citenamefont {Caffarel}(2019)}]{Caffarel_2019}%
  \BibitemOpen
  \bibfield  {author} {\bibinfo {author} {\bibfnamefont {M.}~\bibnamefont
  {Caffarel}},\ }\href {\doibase 10.1063/1.5114703} {\bibfield  {journal}
  {\bibinfo  {journal} {J. Chem. Phys.}\ }\textbf {\bibinfo {volume} {151}},\
  \bibinfo {pages} {064101} (\bibinfo {year} {2019})}\BibitemShut {NoStop}%
\bibitem [{Note1()}]{Note1}%
  \BibitemOpen
  \bibinfo {note} {The property results from the fact that the series is a
  telescoping series and that the general term $\mel { I }{ \qty (T^+_{I})^{n}
  }{ \Psi ^{+}}$ goes to zero as $n$ goes to infinity.}\BibitemShut {Stop}%
\bibitem [{Note2()}]{Note2}%
  \BibitemOpen
  \bibinfo {note} {As $\tau \to 0$ and $N \to \infty $ with $N\tau =t$, the
  operator $T^N$ converges to $e^{-t(H-E \protect \mathds {1})}$. We then have
  $G^E_{ij} \to \DOTSI \intop \ilimits@ _0^{\infty } dt \mel {i}{e^{-t(H-E
  \protect \mathds {1})}}{j}$, which is the Laplace transform of the
  time-dependent Green's function $\mel {i}{e^{-t(H-E \protect \mathds
  {1})}}{j}$.}\BibitemShut {Stop}%
\bibitem [{\citenamefont {Gutzwiller}(1963)}]{Gutzwiller_1963}%
  \BibitemOpen
  \bibfield  {author} {\bibinfo {author} {\bibfnamefont {M.~C.}\ \bibnamefont
  {Gutzwiller}},\ }\href {\doibase 10.1103/PhysRevLett.10.159} {\bibfield
  {journal} {\bibinfo  {journal} {Phys. Rev. Lett.}\ }\textbf {\bibinfo
  {volume} {10}},\ \bibinfo {pages} {159} (\bibinfo {year} {1963})}\BibitemShut
  {NoStop}%
\bibitem [{\citenamefont {Huron}\ \emph {et~al.}(1973)\citenamefont {Huron},
  \citenamefont {Malrieu},\ and\ \citenamefont {Rancurel}}]{Huron_1973}%
  \BibitemOpen
  \bibfield  {author} {\bibinfo {author} {\bibfnamefont {B.}~\bibnamefont
  {Huron}}, \bibinfo {author} {\bibfnamefont {J.~P.}\ \bibnamefont {Malrieu}},
  \ and\ \bibinfo {author} {\bibfnamefont {P.}~\bibnamefont {Rancurel}},\
  }\href {\doibase 10.1063/1.1679199} {\bibfield  {journal} {\bibinfo
  {journal} {J. Chem. Phys.}\ }\textbf {\bibinfo {volume} {58}},\ \bibinfo
  {pages} {5745} (\bibinfo {year} {1973})}\BibitemShut {NoStop}%
\bibitem [{\citenamefont {Harrison}(1991)}]{Harrison_1991}%
  \BibitemOpen
  \bibfield  {author} {\bibinfo {author} {\bibfnamefont {R.~J.}\ \bibnamefont
  {Harrison}},\ }\href {\doibase 10.1063/1.460537} {\bibfield  {journal}
  {\bibinfo  {journal} {J. Chem. Phys.}\ }\textbf {\bibinfo {volume} {94}},\
  \bibinfo {pages} {5021} (\bibinfo {year} {1991})}\BibitemShut {NoStop}%
\bibitem [{\citenamefont {Giner}\ \emph {et~al.}(2013)\citenamefont {Giner},
  \citenamefont {Scemama},\ and\ \citenamefont {Caffarel}}]{Giner_2013}%
  \BibitemOpen
  \bibfield  {author} {\bibinfo {author} {\bibfnamefont {E.}~\bibnamefont
  {Giner}}, \bibinfo {author} {\bibfnamefont {A.}~\bibnamefont {Scemama}}, \
  and\ \bibinfo {author} {\bibfnamefont {M.}~\bibnamefont {Caffarel}},\ }\href
  {\doibase 10.1139/cjc-2013-0017} {\bibfield  {journal} {\bibinfo  {journal}
  {Can. J. Chem.}\ }\textbf {\bibinfo {volume} {91}},\ \bibinfo {pages} {879}
  (\bibinfo {year} {2013})}\BibitemShut {NoStop}%
\bibitem [{\citenamefont {Holmes}\ \emph {et~al.}(2016)\citenamefont {Holmes},
  \citenamefont {Tubman},\ and\ \citenamefont {Umrigar}}]{Holmes_2016}%
  \BibitemOpen
  \bibfield  {author} {\bibinfo {author} {\bibfnamefont {A.~A.}\ \bibnamefont
  {Holmes}}, \bibinfo {author} {\bibfnamefont {N.~M.}\ \bibnamefont {Tubman}},
  \ and\ \bibinfo {author} {\bibfnamefont {C.~J.}\ \bibnamefont {Umrigar}},\
  }\href {\doibase 10.1021/acs.jctc.6b00407} {\bibfield  {journal} {\bibinfo
  {journal} {J. Chem. Theory Comput.}\ }\textbf {\bibinfo {volume} {12}},\
  \bibinfo {pages} {3674} (\bibinfo {year} {2016})}\BibitemShut {NoStop}%
\bibitem [{\citenamefont {Schriber}\ and\ \citenamefont
  {Evangelista}(2016)}]{Schriber_2016}%
  \BibitemOpen
  \bibfield  {author} {\bibinfo {author} {\bibfnamefont {J.~B.}\ \bibnamefont
  {Schriber}}\ and\ \bibinfo {author} {\bibfnamefont {F.~A.}\ \bibnamefont
  {Evangelista}},\ }\href {\doibase 10.1063/1.4948308} {\bibfield  {journal}
  {\bibinfo  {journal} {J. Chem. Phys.}\ }\textbf {\bibinfo {volume} {144}},\
  \bibinfo {pages} {161106} (\bibinfo {year} {2016})}\BibitemShut {NoStop}%
\bibitem [{\citenamefont {Tubman}\ \emph {et~al.}(2020)\citenamefont {Tubman},
  \citenamefont {Freeman}, \citenamefont {Levine}, \citenamefont {Hait},
  \citenamefont {Head-Gordon},\ and\ \citenamefont {Whaley}}]{Tubman_2020}%
  \BibitemOpen
  \bibfield  {author} {\bibinfo {author} {\bibfnamefont {N.~M.}\ \bibnamefont
  {Tubman}}, \bibinfo {author} {\bibfnamefont {C.~D.}\ \bibnamefont {Freeman}},
  \bibinfo {author} {\bibfnamefont {D.~S.}\ \bibnamefont {Levine}}, \bibinfo
  {author} {\bibfnamefont {D.}~\bibnamefont {Hait}}, \bibinfo {author}
  {\bibfnamefont {M.}~\bibnamefont {Head-Gordon}}, \ and\ \bibinfo {author}
  {\bibfnamefont {K.~B.}\ \bibnamefont {Whaley}},\ }\href {\doibase
  10.1021/acs.jctc.8b00536} {\bibfield  {journal} {\bibinfo  {journal} {J.
  Chem. Theory Comput.}\ }\textbf {\bibinfo {volume} {16}},\ \bibinfo {pages}
  {2139} (\bibinfo {year} {2020})}\BibitemShut {NoStop}%
\bibitem [{\citenamefont {Booth}\ \emph {et~al.}(2009)\citenamefont {Booth},
  \citenamefont {Thom},\ and\ \citenamefont {Alavi}}]{Booth_2009}%
  \BibitemOpen
  \bibfield  {author} {\bibinfo {author} {\bibfnamefont {G.~H.}\ \bibnamefont
  {Booth}}, \bibinfo {author} {\bibfnamefont {A.~J.~W.}\ \bibnamefont {Thom}},
  \ and\ \bibinfo {author} {\bibfnamefont {A.}~\bibnamefont {Alavi}},\ }\href
  {\doibase 10.1063/1.3193710} {\bibfield  {journal} {\bibinfo  {journal} {J.
  Chem. Phys.}\ }\textbf {\bibinfo {volume} {131}},\ \bibinfo {pages} {054106}
  (\bibinfo {year} {2009})}\BibitemShut {NoStop}%
\bibitem [{\citenamefont {Cleland}\ \emph {et~al.}(2010)\citenamefont
  {Cleland}, \citenamefont {Booth},\ and\ \citenamefont
  {Alavi}}]{Cleland_2010}%
  \BibitemOpen
  \bibfield  {author} {\bibinfo {author} {\bibfnamefont {D.}~\bibnamefont
  {Cleland}}, \bibinfo {author} {\bibfnamefont {G.~H.}\ \bibnamefont {Booth}},
  \ and\ \bibinfo {author} {\bibfnamefont {A.}~\bibnamefont {Alavi}},\ }\href
  {\doibase 10.1063/1.3302277} {\bibfield  {journal} {\bibinfo  {journal} {J.
  Chem. Phys.}\ }\textbf {\bibinfo {volume} {132}},\ \bibinfo {pages} {041103}
  (\bibinfo {year} {2010})}\BibitemShut {NoStop}%
\bibitem [{\citenamefont {Zhang}\ and\ \citenamefont
  {Krakauer}(2003)}]{Zhang_2003}%
  \BibitemOpen
  \bibfield  {author} {\bibinfo {author} {\bibfnamefont {S.}~\bibnamefont
  {Zhang}}\ and\ \bibinfo {author} {\bibfnamefont {H.}~\bibnamefont
  {Krakauer}},\ }\href {\doibase 10.1103/PhysRevLett.90.136401} {\bibfield
  {journal} {\bibinfo  {journal} {Phys. Rev. Lett.}\ }\textbf {\bibinfo
  {volume} {90}},\ \bibinfo {pages} {136401} (\bibinfo {year}
  {2003})}\BibitemShut {NoStop}%
\bibitem [{\citenamefont {White}\ and\ \citenamefont
  {Martin}(1999)}]{White_1999}%
  \BibitemOpen
  \bibfield  {author} {\bibinfo {author} {\bibfnamefont {S.~R.}\ \bibnamefont
  {White}}\ and\ \bibinfo {author} {\bibfnamefont {R.~L.}\ \bibnamefont
  {Martin}},\ }\href {\doibase 10.1063/1.478295} {\bibfield  {journal}
  {\bibinfo  {journal} {J. Chem. Phys.}\ }\textbf {\bibinfo {volume} {110}},\
  \bibinfo {pages} {4127} (\bibinfo {year} {1999})}\BibitemShut {NoStop}%
\bibitem [{\citenamefont {Chan}\ and\ \citenamefont
  {Sharma}(2011)}]{Chan_2011}%
  \BibitemOpen
  \bibfield  {author} {\bibinfo {author} {\bibfnamefont {G.~K.-L.}\
  \bibnamefont {Chan}}\ and\ \bibinfo {author} {\bibfnamefont {S.}~\bibnamefont
  {Sharma}},\ }\href {\doibase 10.1146/annurev-physchem-032210-103338}
  {\bibfield  {journal} {\bibinfo  {journal} {Annu. Rev. Phys. Chem.}\ }\textbf
  {\bibinfo {volume} {62}},\ \bibinfo {pages} {465} (\bibinfo {year}
  {2011})}\BibitemShut {NoStop}%
\end{thebibliography}%
\end{document}